\documentclass[twocolumn,superscriptaddress,nofootinbib]{revtex4}

\usepackage{amsmath}
\usepackage{epsfig}
\usepackage{graphics}
\usepackage{graphicx}
\usepackage[colorlinks=true, citecolor=blue, urlcolor = blue, linkcolor= red, bookmarks=true]{hyperref}

\begin{document}
\title{Shadow of rotating charged black hole with Weyl corrections}

\author{Md Sabir Ali } 
\email{alimd.sabir3@gmail.com}
\affiliation{Center for Theoretical Physics, Jamia Millia Islamia, New Delhi 110025, India}

\author{Muhammed Amir}
\email{amirctp12@gmail.com}
\affiliation{Astrophysics and Cosmology Research Unit, School of Mathematics, 
Statistics and Computer Science, University of KwaZulu-Natal, 
Private Bag X54001, Durban 4000, South Africa}
\date{\today }

\begin{abstract}
We construct theoretical investigation of the black hole shadow for rotating charged black hole in an 
asymptotically flat, axisymmetric, and stationary spacetime with Weyl corrections. This spacetime is 
characterized by mass ($M$), charge parameter ($q$), rotation parameter ($a$), and Weyl coupling constant 
($\alpha$). We derive photon geodesics around the black hole and compute expressions for impact parameters 
with the help of photon spherical orbits conditions. We show how the presence of coupling constant $\alpha$ 
affects the shapes of black hole shadow from the usual Kerr-Newman black hole. A comparison with the standard 
Kerr and Kerr-Newman black hole is also include to observe the potential deviation from them. We find that 
the radius of black hole shadow decreases and the distortion in the shape of shadow increases with an 
increase in charge $q$ for both positive and negative values of coupling constant $\alpha$. We further extend 
our study by considering the plasma environment around the black hole and find out the essential expressions 
for the black hole shadow.
\end{abstract}

\maketitle

\section{INTRODUCTION}
\label{intro}
The existence of black hole binaries has been confirmed with the detection of gravitational waves by LIGO 
and VIRGO observations \cite{Abbott:2016blz}. There becomes a common belief after the outstanding 
discovery of quasar \cite{Schmidt:1963} that the supermassive black holes exist at the galactic center, 
and astrophysical black hole candidates are the compact objects in X-ray binary systems 
\cite{Rees:1984si,Kormendy:1995}. These supermassive black holes are contemplating 
as the astrophysical black holes and their formation could be considered as a consequence of the 
gravitational collapse of the matter. The observational evidence in our galaxy support the 
existence of supermassive black hole candidate, Sagittarius A${^*}$ (Sgr A${^*}$) \cite{Eckart:1996}. 
Apart from Sgr A${^*}$, another supermassive black hole is M87 $^*$ which located at the centre of giant 
elliptical galaxy M87 \cite{Broderick:2015tda}.
To analyze the existence and actual nature of black holes demands the direct detection of its event horizon, 
a smooth boundary from where nothing can escape including light. The event horizon always plays a central 
role while discussing the direct observation of the black hole. Although black holes are the strong 
gravitational field regime and the gravitational lensing is the most common phenomenon around them. 
Consequently, outside the event horizon, there exist photon captured regions where photons follow 
circular orbits. We wish to observe these unstable photon orbits which can be considered as the shadow images
of the black hole on the luminous background \cite{Bardeen:1973gb,Chandrasekhar:1992}. Therefore, 
observing the black hole shadow is one of the most outstanding research areas which provides a 
tentative way towards the direct observation of the black holes. Supermassive black hole candidate 
Sgr A${^*}$ is situated at a very large distance ($\sim$8.3kpc) from the earth and its diameter is very tiny 
($\sim$50$\mu$as), which can only be resolved through Very Long Baseline Interferometry (VLBI) observations 
on the sub-millimeter scale with the help of Event Horizon Telescope (EHT) 
\cite{Doeleman:2008qh,Doeleman:2010,Broderick:2009ph}. Besides this, another scientific project is European 
Council based BlackHoleCam \cite{Goddi:2016jrs}, a subsequent member of EHT which has been actively 
participated in the observation to investigate the physics of black hole candidate Sgr A${^*}$ and M87$^*$. 
Recently, the EHT team has announced the first ever picture of supermassive black hole M87$^*$ 
\cite{Akiyama:2019cqa,Akiyama:2019fyp,Akiyama:2019eap}, which is the most outstanding result. There next 
goal is to get the picture of supermassive black hole Sgr A${^*}$. Through such investigation, we will 
be able to discuss the fundamental properties of the black holes. The study of shadow templates can be 
expected to have strong evidences for the existence of stationary black holes in theories other than 
Einstein's general relativity.

A wide range of research articles is available regarding the theoretical study of the black hole shadow. 
The theoretical investigations suggest that the shape of black hole shadow can give us the paramount 
information about the parameters of particular black hole spacetime. The study on the shadow of the Kerr 
black hole described that the shape is very sensitive to the spin; it distorted the shape and the distortion 
increased with an increase in spin \cite{Hioki:2009na}. There exists various theoretical methods for the 
characterization of black hole shadow \cite{Hioki:2009na,Abdujabbarov:2015xqa,Younsi:2016azx}. However, 
the shadow has been discussed for a wide range of different black hole spacetimes in four dimensions 
\cite{Takahashi:2005hy,Bambi:2008jg,Wei:2013kza,Bambi:2010hf,Amarilla:2010zq,Amarilla:2011fx,Amarilla:2013sj,Yumoto:2012kz,Abdujabbarov:2012bn,Atamurotov:2013sca,Li:2013jra,Grenzebach:2014fha,Cunha:2015yba,Johannsen:2015qca,Abdujabbarov:2016hnw,Atamurotov:2015nra,Amir:2016cen,Kumar:2017vuh} 
as well as in higher dimensions \cite{Papnoi:2014aaa,Amir:2017slq,Singh:2017vfr}. Apart from the black holes, 
the theoretical investigation of shadow for other compact objects, e.g., wormholes have also been discussed in
\cite{Nedkova:2013msa,Ohgami:2015nra,Abdujabbarov:2016efm,Shaikh:2018kfv,Amir:2018szm,Amir:2018pcu}.

In the current work, we consider a special spacetime, namely rotating charged black holes with Weyl 
corrections \cite{Chen:2013raa} to construct black hole shadow. The spacetime comprises of coupling in 
between the Weyl tensor and the Maxwell field \cite{Chen:2013raa}. In other words, it can be considered as 
a coupling between the electromagnetic field and the gravitational field. Interestingly, this coupling has 
an astrophysical significance: the authors in \cite{Preuss:2004pp,Solanki:2004az} investigated that such a 
coupling could exist near highly massive astrophysical compact objects. Taking into account such couplings, 
the other interesting phenomena has also been discussed in literature, e.g., holographic superconductors 
\cite{Wu:2010vr,Ma:2011zze,Roychowdhury:2012hp}, holographic conductivity \cite{Ritz:2008kh} etc.
Since the astrophysical black holes are expected to be described by the Kerr spacetime, hence one cannot 
avoid the possibility of black hole solution in other theories of gravity. It will be very intersting and 
helpful to test these theories by observing the deviations from the Kerr spacetime and extract information 
about the deviation parameters as well other parameters that lie within the particular theory. 
Therefore, it is necessary step to develop theoretical investigations of the shadow for different 
black hole spacetimes and compare the results with the Kerr and Kerr-Newman spacetimes. This motivates 
us toward the investigation of the shadow of rotating charged black holes with Weyl corrections and we 
explore how the presence of Weyl corrections in the spacetime affect the shape of the shadow. 

The outlines of our paper is as follows. We briefly review the rotating charged black holes with Weyl 
corrections in Sec.~\ref{spacetime}. Photon geodesics around the black hole is the subject of 
Sec.~\ref{geod}. In Sec.~\ref{shad}, we investigate the shadow of the rotating charged black hole with Weyl 
corrections and construct various images for different cases. In Sec.~\ref{plasma}, we examine the 
formulation of the shadow in surrounding plasma environment. We end by concluding our results in 
Sec.~\ref{concl}. We have chosen ($-,+,+,+$) signature convention and geometrized units ($G=c=1$) 
throughout the paper.

\section{Rotating charged black hole with Weyl corrections}
\label{spacetime}
We briefly discuss about the rotating charged black hole with Weyl corrections. The charged black holes with 
Weyl corrections and their rotating counterpart have been  investigated by Songbai Chen \emph{et al.} in 
\cite{Chen:2013raa}. The action that represents the coupling of electromagnetic field and Weyl tensor is 
proposed \cite{Chen:2013raa} as
\begin{eqnarray}\label{action}
S = \int d^4 x \sqrt{-g} \left[R - \frac{1}{4} F_{\mu \nu} F^{\mu \nu} +\alpha C^{\mu \nu \rho \sigma}
F_{\mu \nu} F^{\mu \nu}\right],
\end{eqnarray}
where $F_{\mu \nu} = \partial_{\mu}A_{\nu} -\partial_{\nu}A_{\mu}$ is the electromagnetic tensor and 
$A_{\mu}$ is the vector potential have the following form
\begin{equation}
A_{\mu} = \left(\phi(r), 0, 0, 0\right).
\end{equation}
As a consequence the spherically symmetric spacetime of the charged black hole with Weyl corrections 
is obtained
\cite{Chen:2013raa} as follows
\begin{eqnarray}
ds^2 &=& -\chi(r) dt^2 + \frac{1}{\chi(r)} dr^2 \nonumber\\ 
&& + \left(r^2 +\frac{4\alpha q^2}{9 r^2}\right) 
\left(d \theta^2 + \sin^2 \theta d \varphi^2 \right),
\end{eqnarray}
where the metric function $\chi (r)$ and the static electric potential $\phi (r)$ are given by
\begin{eqnarray}\label{ci}
\chi(r) &=& 1-\frac{2M}{r}+\frac{q^2}{r^2}-\frac{4 \alpha q^2}{3 r^4}
\left(1 -\frac{10 M}{3r} + \frac{26q^2}{15 r^2}\right), \nonumber\\
\phi(r) &=& \frac{q}{r} +\frac{\alpha q}{r^3} \left(\frac{M}{r} -\frac{37 q^2}{45 r^2} \right).
\end{eqnarray}
It can be seen that the electric potential $\phi(r)$ has an additional dependency on mass $M$ and $\alpha$ 
when we compare it with the Reissner-Noredstr{\"o}m black hole. Therefore, it deviates from the standard 
Reissner-Nordstr{\"o}m spacetime. The spacetime metric of the rotating charged black hole with Weyl corrections \cite{Chen:2013raa} when applying the Newman-Janis Algorithm, in Boyer-Lindquist $(t, r, \theta, \phi)$ coordinates reads
\begin{eqnarray}\label{metric}
ds^2 &=& -\chi(r,\theta)dt^2  +\frac{\tilde{\Sigma}}{\Delta}dr^2  \nonumber\\
&& -2 a\left(1-\chi(r,\theta)\right) \sin^2 \theta dt d\varphi + \tilde{\Sigma} d\theta^2 \nonumber\\
&& +\sin^2\theta \left[\tilde{\Sigma} +a^2\left(2-\chi(r,\theta)\right)
\sin^2\theta\right]d \varphi^2,
\end{eqnarray}
where the metric function $\chi(r,\theta)$ is 
\begin{eqnarray}\label{metfunc}
\chi(r,\theta) = 1-\frac{2Mr}{\Sigma}-\frac{q^2}{\Sigma}-\frac{4 \alpha q^2}{3 \Sigma^2}
\left(1-\frac{50Mr-26q^2}{15 \Sigma}\right).
\end{eqnarray}
Here $\Sigma = r^2+a^2\cos^2\theta$ and the other unknown terms in (\ref{metric}) are expressed as follows
\begin{eqnarray}\label{func}
\tilde{\Sigma} = \Sigma +4 \alpha q^2/9 \Sigma, \quad 
\Delta = \tilde{\Sigma} \chi(r,\theta) +a^2\sin^2\theta,
\end{eqnarray}
where $M$ is the black hole mass, $a=J/M$ is the rotation parameter where $J$ is angular momentum of the black 
hole, $q$ is the electric charge, and $\alpha$ is the coupling constant having dimension of length squared. The contravariant components of the metric (\ref{metric}), are calculated to be
\begin{eqnarray}
g^{tt} &=& -\frac{\left(\tilde{\Sigma} +a^2\sin^2\theta\right)^2-\Delta a^2\sin^2\theta}
{\tilde{\Sigma} \Delta}, \nonumber\\
g^{t \varphi} &=& -\frac{a(\tilde{\Sigma} +a^2\sin^2\theta-\Delta)}{\tilde{\Sigma} \Delta}, \nonumber\\
g^{rr} &=& \frac{\Delta}{\tilde{\Sigma}}, \quad
g^{\theta\theta} = \frac{1}{\tilde{\Sigma}}, \nonumber\\
g^{\varphi \varphi} &=& \frac{\Delta -a^2 \sin^2\theta}{\tilde{\Sigma} \Delta \sin^2 \theta}.
\end{eqnarray}
The spacetime metric (\ref{metric}) are affected by coupling parameter $\alpha$, which for $\alpha=0$ and $\alpha=0=q$, 
respectively, reduces to Kerr-Newman and the Kerr metric. If one substitutes $a=0$, then it reduces to the charged 
spherically symmetric metric with Weyl corrected term. When $q=a=0$, the metric (\ref{metric}) reduces to the Schwarzschild 
metric. The metric (\ref{metric}) contains two roots corresponding to event horizon and Cauchy horizon, degenerate roots 
when two horizons coincide, and a naked singularity when no solutions exist for a range of values of the parameters. 
A detailed analysis of the horizons and static limit surface has been discussed in \cite{Chen:2013raa}. In the next 
section, we shall discuss the photons motion in the gravitational field of rotating charged black hole with Weyl corrections.

\section{Photon geodesics}
\label{geod}
In order to construct the formalism for the black hole shadow, we need to calculate the photon geodesics or 
null geodesics in the gravitational field of the rotating charged black hole with Weyl corrections. Since the 
spacetime metric (\ref{metric}) contains two Killing vectors, namely $\xi^{\mu}_{(t)}$ and 
$\xi^{\mu}_{(\varphi)}$ corresponding to time-translation symmetry along $t$-axis and rotational symmetry 
about $\varphi$-axis. Now we write down the Lagrangian for the metric
\begin{eqnarray}\label{lag}
\mathcal{L} &=& \frac{1}{2} g_{\mu \nu} u^{\mu} u^{\nu}, \nonumber\\
&=& \frac{1}{2}\left(g_{tt} \dot{t}^2+2g_{t\varphi} \dot{t}\dot{\varphi} +g_{rr} \dot{r}^2 
+g_{\theta\theta} \dot{\theta}^2+g_{\varphi \varphi}\,\dot{\varphi}^2\right),
\end{eqnarray}
where $u^{\mu}=d x^{\mu}/d \sigma = \dot{x}^{\mu}$ ($\mu = 0,1,2,3$) is the four-velocity of the photon. We 
analyse that the Lagrangian (\ref{lag}) does not have any dependence on coordinates $t$ and $\varphi$. 
Therefore, the conserved quantities associated to the  Killing vectors $\xi^{\mu}_{(t)}$ and 
$\xi^{\mu}_{(\varphi)}$, respectively, are the energy $E=-\xi^{\mu}_{(t)}u_{\mu}$ and the angular momentum 
$ L_z =\xi^{\mu}_{(\varphi)}u_{\mu}$, which can be easily computed to give
\begin{eqnarray}\label{cons}
-E = p_t = g_{tt}\dot{t}+g_{t\varphi}\dot{\varphi}, \quad 
L_z = p_\varphi = g_{t\varphi}\dot{t}+g_{\varphi \varphi}\dot{\varphi},
\end{eqnarray}
where $p_{\mu}= \partial \mathcal{L}/\partial \dot{x_{\mu}}$ is the four-momentum of the photon. On solving 
the Eq.~(\ref{cons}) simultaneously, for $\dot{t}$ and $\dot{\varphi}$, we obtain the four-velocities
\begin{eqnarray}\label{t-phi}
\dot{t} = \frac{g_{\varphi\varphi}\,E+g_{t\varphi}\,L_z}{g_{t\varphi}^2-g_{tt}g_{\varphi \varphi}}, \quad
\dot{\varphi} =-\frac{g_{t\varphi}\,E+g_{tt}\,L_z}{g_{t\varphi}^2-g_{tt}g_{\varphi \varphi}}.
\end{eqnarray}
As can be seen from metric~(\ref{metric}) that the functions have very complicated forms, therefore it is 
impossible to apply the variable separation method using the Hamilton-Jacobi formulation. In order 
to resolve this ambiguity, we use the approximation in the polar coordinate $\theta$ such that 
$\theta \approx \pi/2+\epsilon$, where $\epsilon$ is the very tiny angle deviated from equatorial plane. 
Within this approximation, the trigonometric functions transform into $\sin\theta=1$, 
$\cos\theta=\epsilon$, and the function $\chi(r,\theta)$ takes the form of Eq.~(\ref{ci}).
Henceforth, we shall use the notation $\chi$ instead of $\chi(r)$ throughout the paper. It is noticeable
that we are considering the photon trajectories very close to the equatorial plane. Since the observer is 
situated at infinity, therefore, the photons will approach to the region very close to the equatorial plane.
However, the fact that unstable photon spherical trajectories are not restricted necessarily on 
equatorial plane; and the consideration in \cite{Abdujabbarov:2016hnw} does not spoil the calculation.
Now by substituting the metric components into the Eq.~(\ref{t-phi}), we obtain the geodesic equations for 
$\dot{t}$ and $\dot{\varphi}$ as follows
\begin{eqnarray}\label{tphi}
r^2 \lambda \frac{d{t}}{d{\sigma}} &=&
a\left(L_z-aE\right)+\frac{\mathcal{P}}{\Delta}\left(r^2+a^2+\frac{4\alpha q^2}{9r^2}\right), 
\nonumber \\
r^2 \lambda \frac{d{\varphi}}{d{\sigma}}&=& \left(L_z-aE\right)+\frac{a \mathcal{P}}{\Delta},
\end{eqnarray}
where $\mathcal{P} :=\left(r^2 +a^2 +\frac{4\alpha q^2}{9r^2}\right)E-aL_z $ and
$\lambda := 1+\frac{4\alpha q^2}{9r^4}$. In order to compute the radial and angular geodesic equations 
for the photon the Hamilton-Jacobi formulation is the best method to apply. The corresponding Hamilton-Jacobi equation 
is given by 
\begin{equation}\label{hj}
\frac{\partial S}{\partial\sigma}=-\frac{1}{2}g^{\mu\nu}\frac{\partial S}{\partial{x^\mu}}
\frac{\partial S}{\partial{x^\nu}}\;,
\end{equation}
where $S$ is the Jacobi action. Although a separable solution for Eq.~(\ref{hj}) can be expressed by 
following \emph{ansatz}
\begin{equation}\label{hj1}  
S=\frac{1}{2}m^2\sigma -Et +L_z\varphi +S_r(r) +S_\epsilon({\epsilon}),
\end{equation}
where $S_{r}(r)$ and $S_{\epsilon}(\epsilon)$ are the functions of only $r$ and $\epsilon$, respectively. 
Now on inserting Eq.~(\ref{hj1}) into Eq.~(\ref{hj}) and separating out the terms of variables $r$ and $\epsilon$, 
eventually equating them to the Carter constant $\mathcal{K}$ which turns out the following geodesic equations
\begin{eqnarray}\label{rtheta}
r^2 \lambda \frac{d{r}}{d{\sigma}}&=&\pm\sqrt{\mathcal{R}}, \nonumber\\
r^2 \lambda \frac{d{\epsilon}}{d{\sigma}}&=&\pm\sqrt{{\Theta}},
\end{eqnarray} 
where
\begin{eqnarray}\label{RT}
\mathcal{R} &=& \mathcal{P}^2-\Delta\left[\mathcal{K}+\left(L_z-aE\right)^2\right], \nonumber\\
\Theta &=& \mathcal{K}.
\end{eqnarray}
The plus $``+"$ and minus $``-"$ signs in Eq.~(\ref{rtheta}) corresponds to the outgoing and the ingoing 
photon trajectories, respectively. These four geodesics equations corresponding to the propagation of light 
in the background of rotating charged black hole with Weyl corrections. Now we redefine our conserved 
quantities in terms of new quantities in order to reduce the number of parameters such that $\xi=L_z/E$ and 
$\eta=\mathcal{K}/E^2$. The conditions for the unstable spherical photon orbits are expressed as follows
\begin{equation}
\label{ucond1}
\mathcal{R}=0, \quad \frac{d\mathcal{R}}{dr}=0. 
\end{equation}
On substituting $\mathcal{R}$ from Eq.~(\ref{RT}) into Eq.~(\ref{ucond1}), and after some straightforward 
computation, we are able to get the expressions of impact parameters $\xi$ and $\eta$ in the following forms
\begin{eqnarray}\label{xiet}
\xi &=& \frac{\left(r^2 \lambda +a^2 \right) \left(\lambda r \chi^{\prime} +2\gamma \chi \right) 
-4\gamma \left(\lambda \chi +a^2\right)}{a\left(r \lambda \chi^{\prime} +2\gamma \chi \right)}, \nonumber\\
\eta &=& \frac{r^3\left[8a^2\gamma \lambda^2 \chi^{\prime} -r\lambda^2 \left(r \chi^{\prime} 
-2\gamma \lambda\chi \right)^2\right]}{a^2 \left(r\lambda \chi^{\prime} +2\gamma \chi \right)^2}.
\end{eqnarray}
Here we define $\gamma :=  1-\frac{4\alpha q^2}{9r^4}$ and $\chi^{\prime}$ is the derivative of the metric 
function $\chi$ with respect to $r$, which is given by
\begin{equation}
\chi^{\prime} =\frac{2\left(Mr-q^2\right)}{r^3}+\frac{16\alpha q^2}{r^5}
\left[1-\frac{125Mr-26q^2}{10r^2}\right].
\end{equation}
The impact parameters which we have obtained in Eq.~(\ref{xiet}) match with the expressions of Kerr-Newman 
black hole when we substitute $\alpha=0$. On the other hand, for $\alpha=q=0$, they reduce to the impact 
parameters of the Kerr black hole. Consequently, we will be able to find out the deviations from the Kerr 
and the Kerr-Newman spacetimes.

\section{Black hole shadow with Weyl corrections}
\label{shad}
\begin{figure}[b]
\includegraphics[scale=0.3]{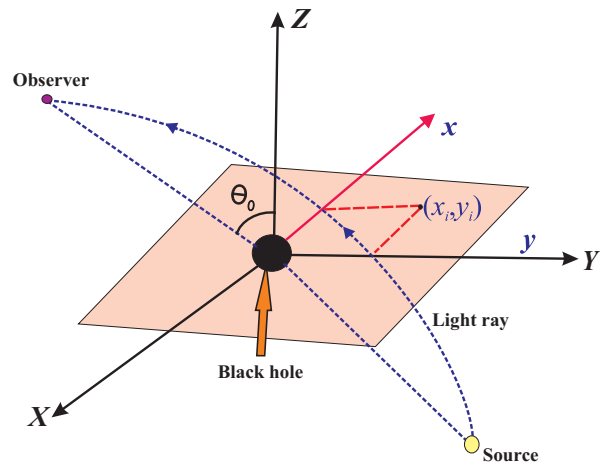}
\caption{\label{celest} Illustration of celestial plane of the observer's sky.}
\end{figure}

\begin{figure*}[t]
\begin{tabular}{c c c c}
\includegraphics[scale=0.35]{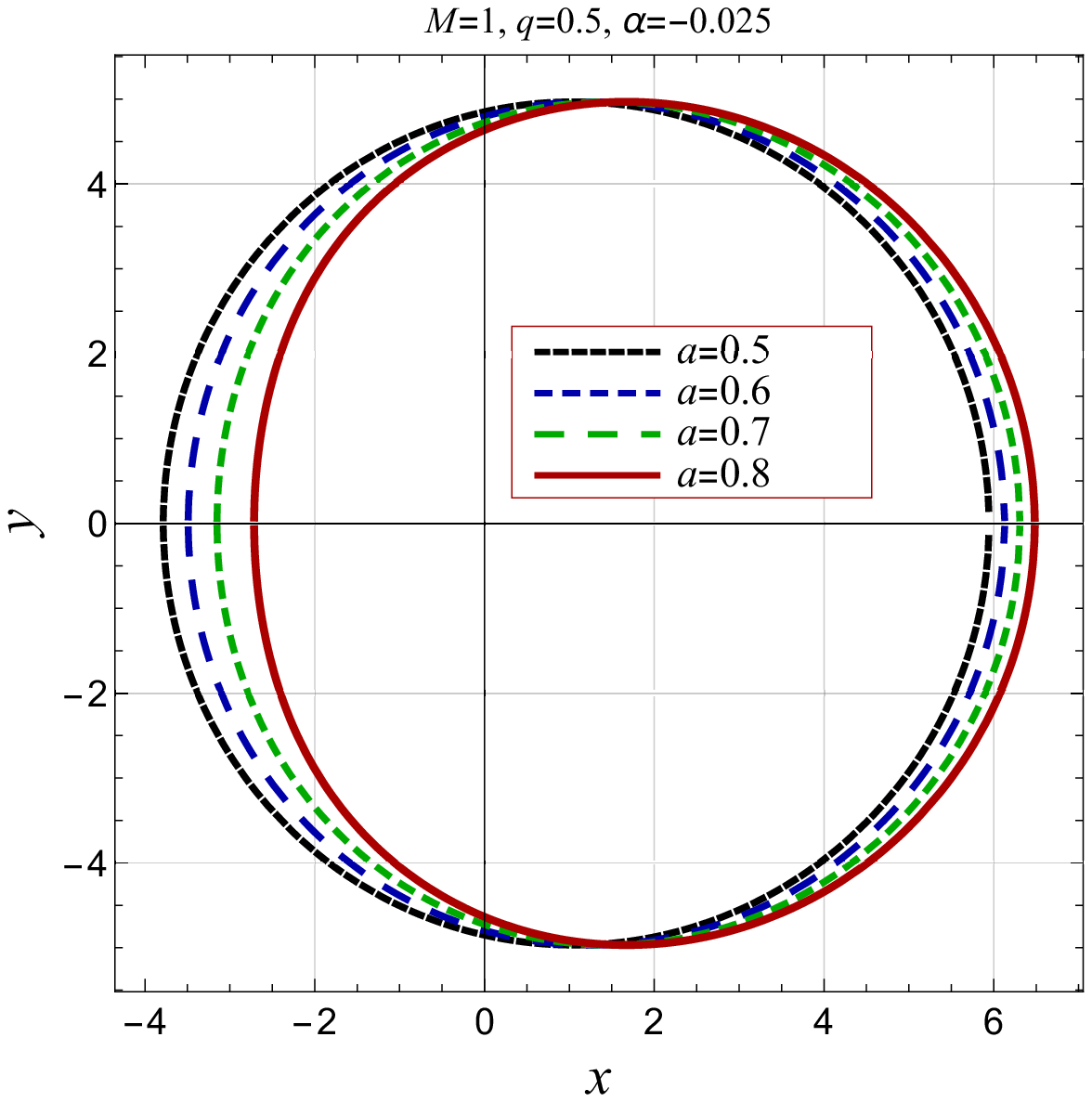}
\includegraphics[scale=0.35]{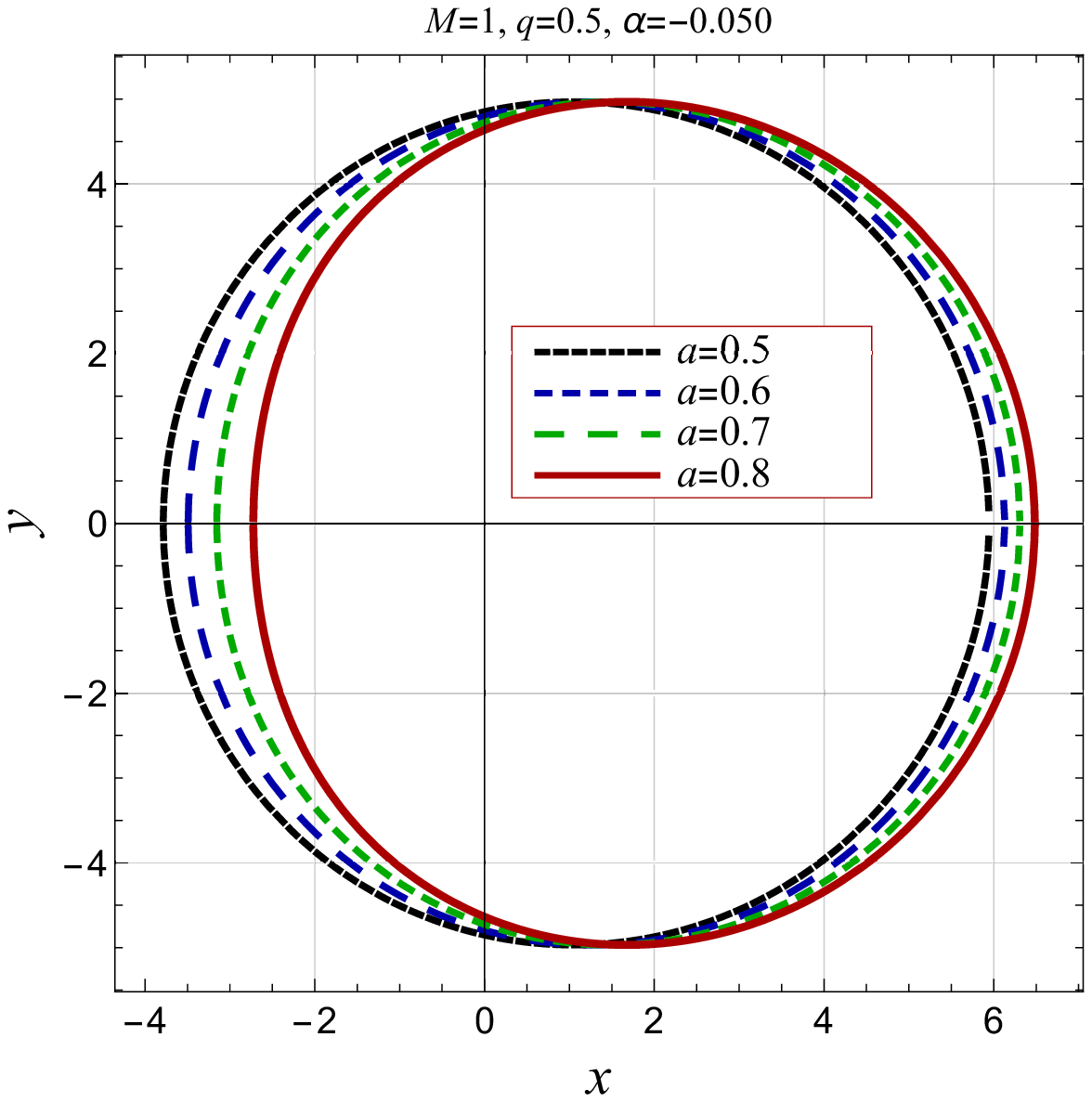}
\includegraphics[scale=0.35]{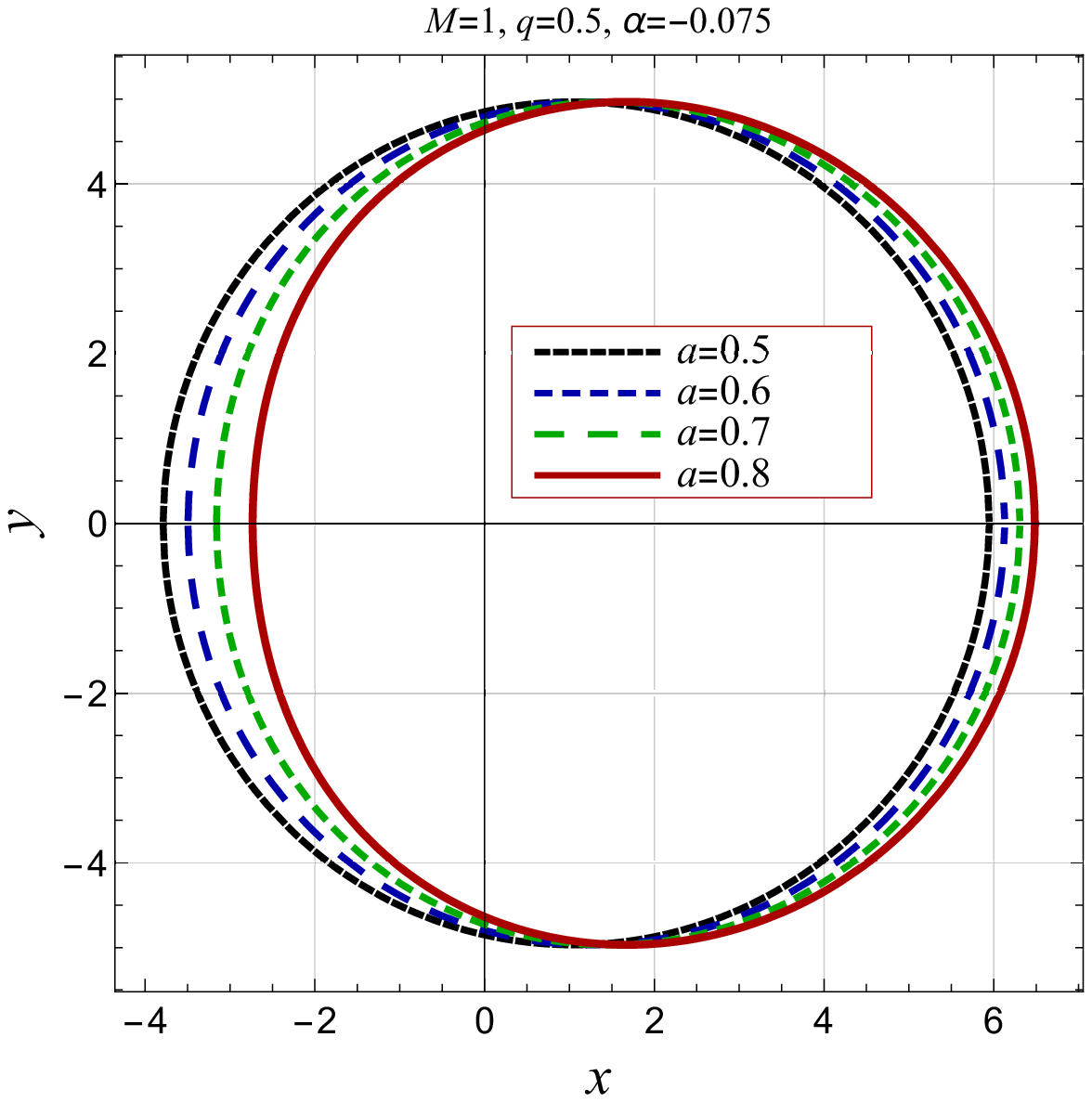}
\includegraphics[scale=0.35]{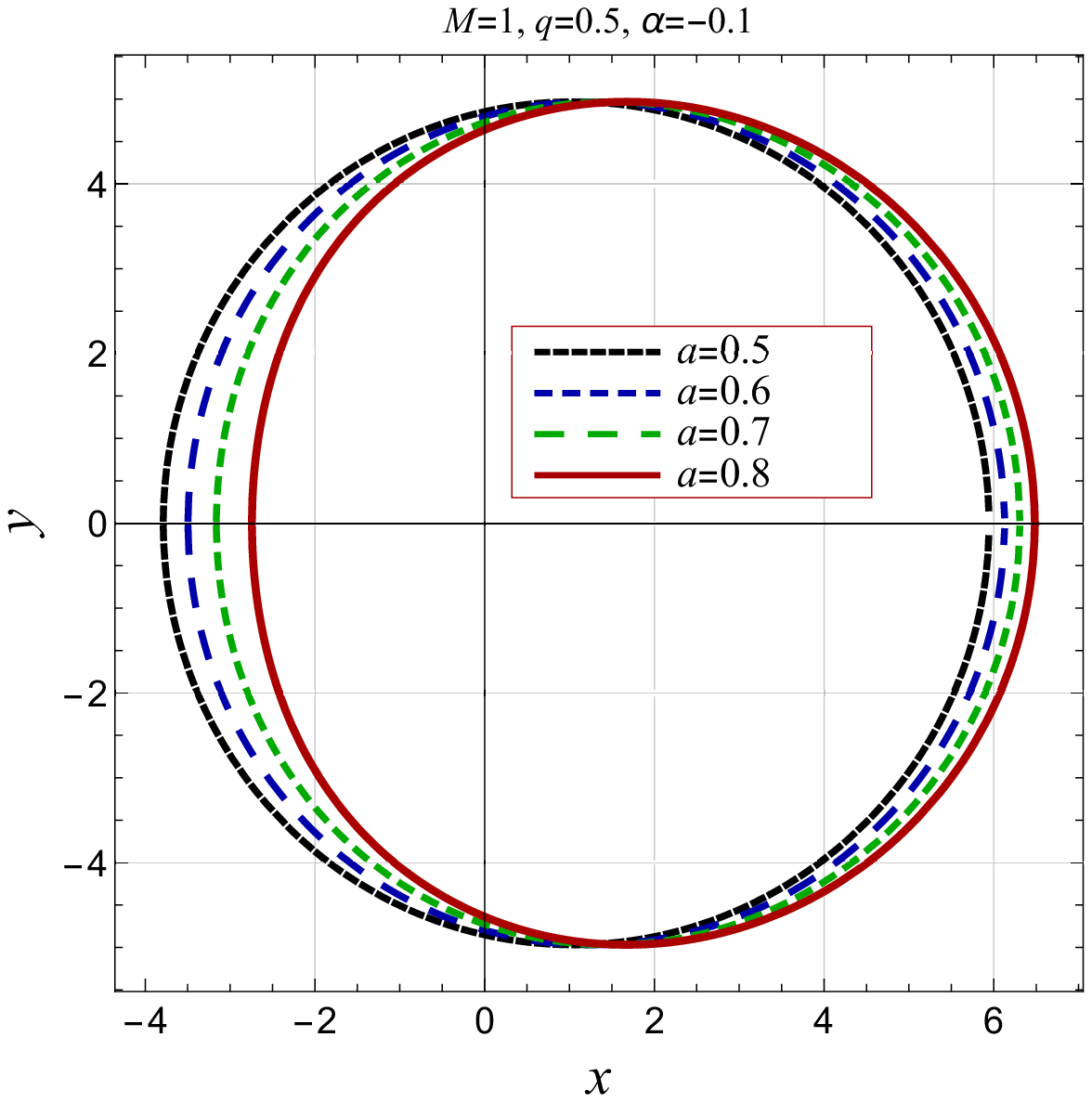}\\
\includegraphics[scale=0.35]{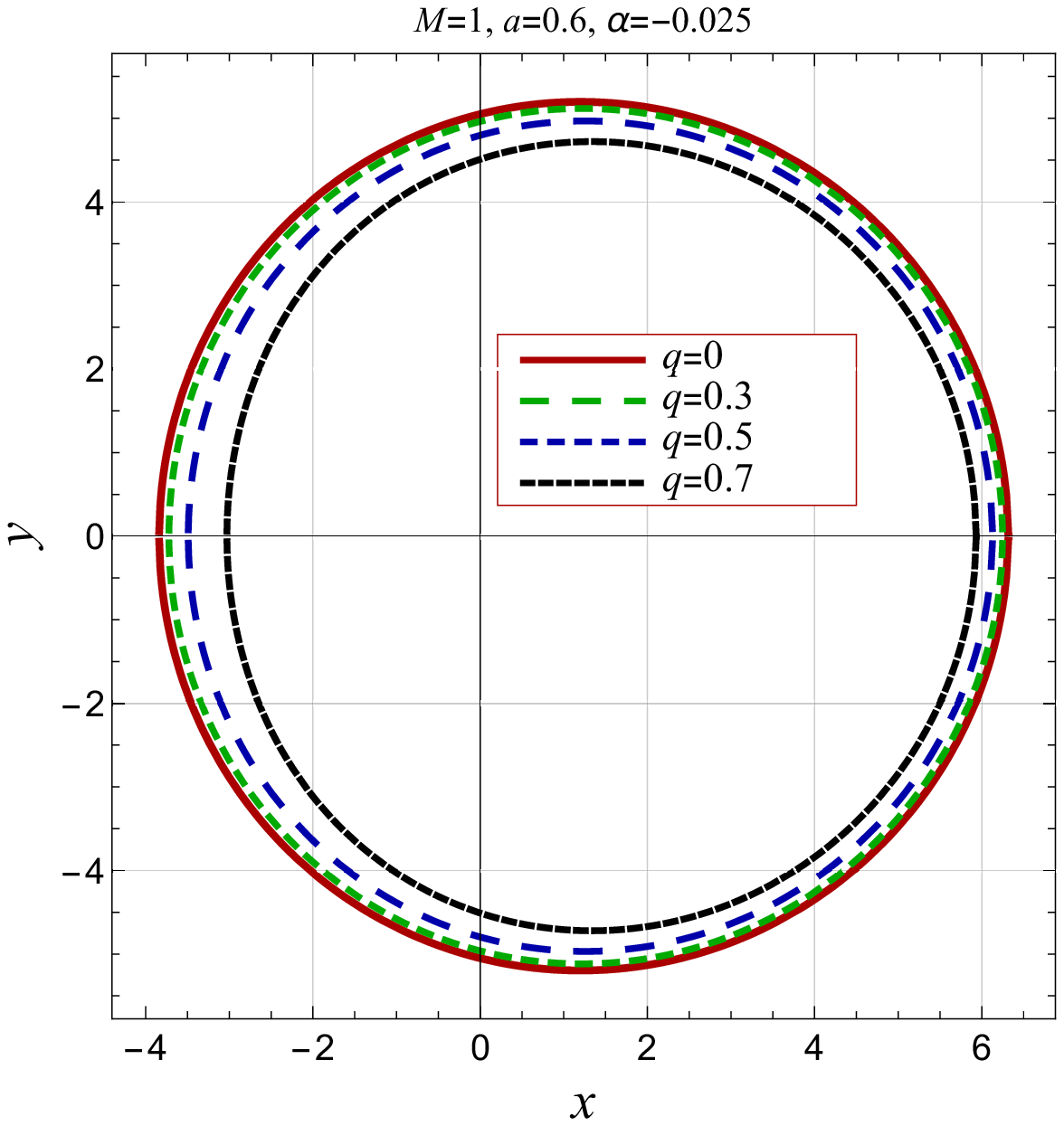}
\includegraphics[scale=0.35]{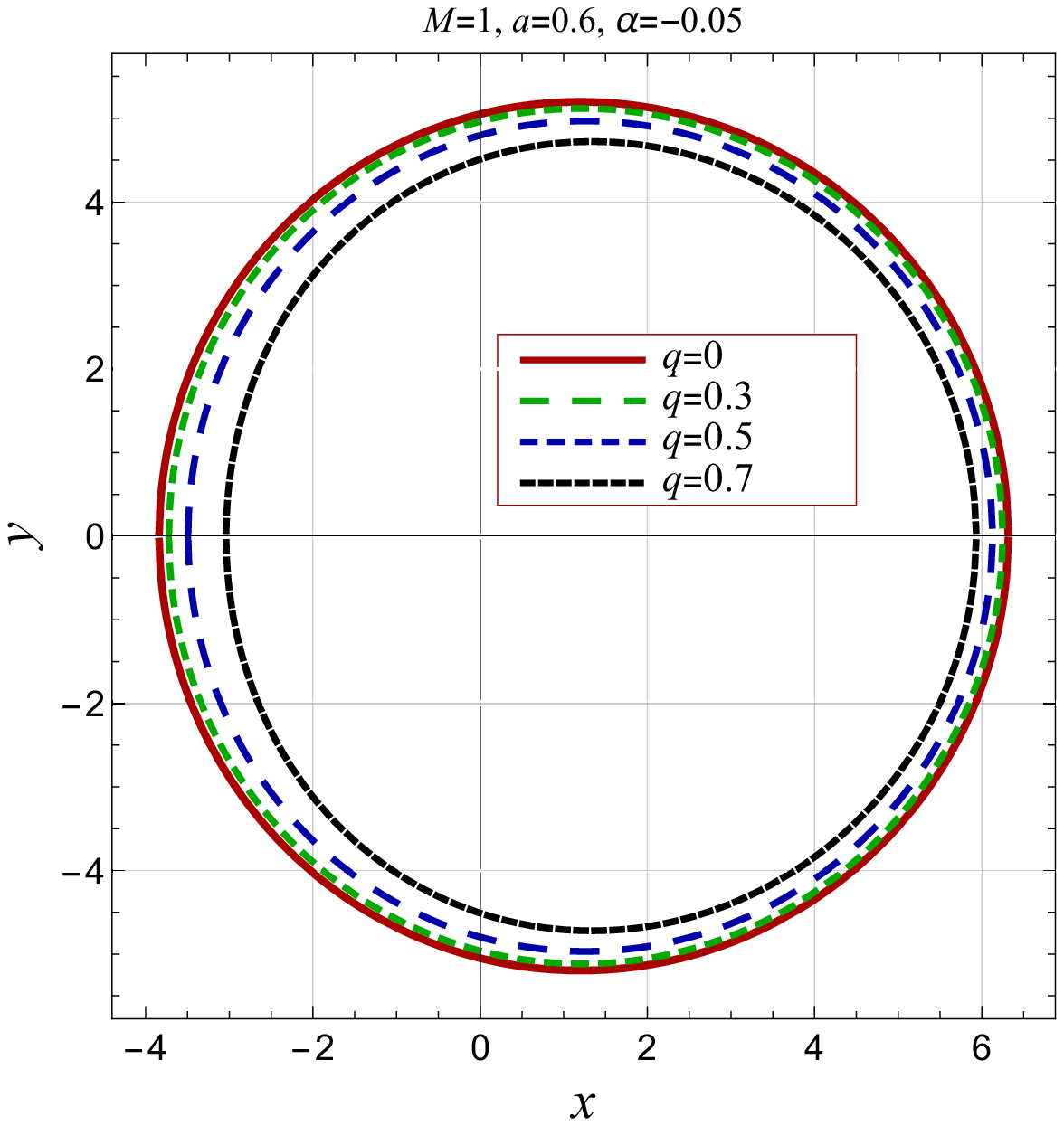}
\includegraphics[scale=0.35]{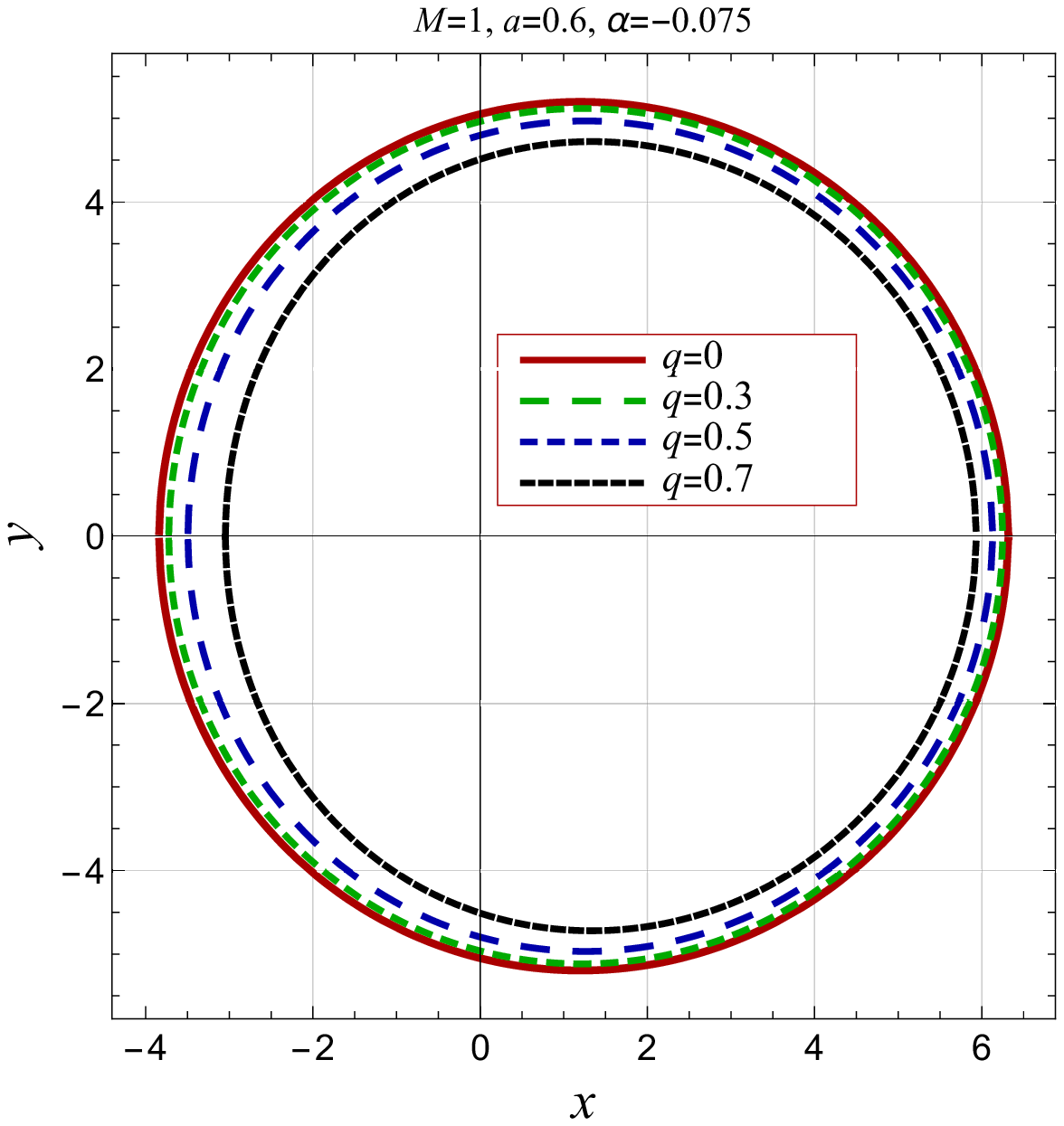}
\includegraphics[scale=0.35]{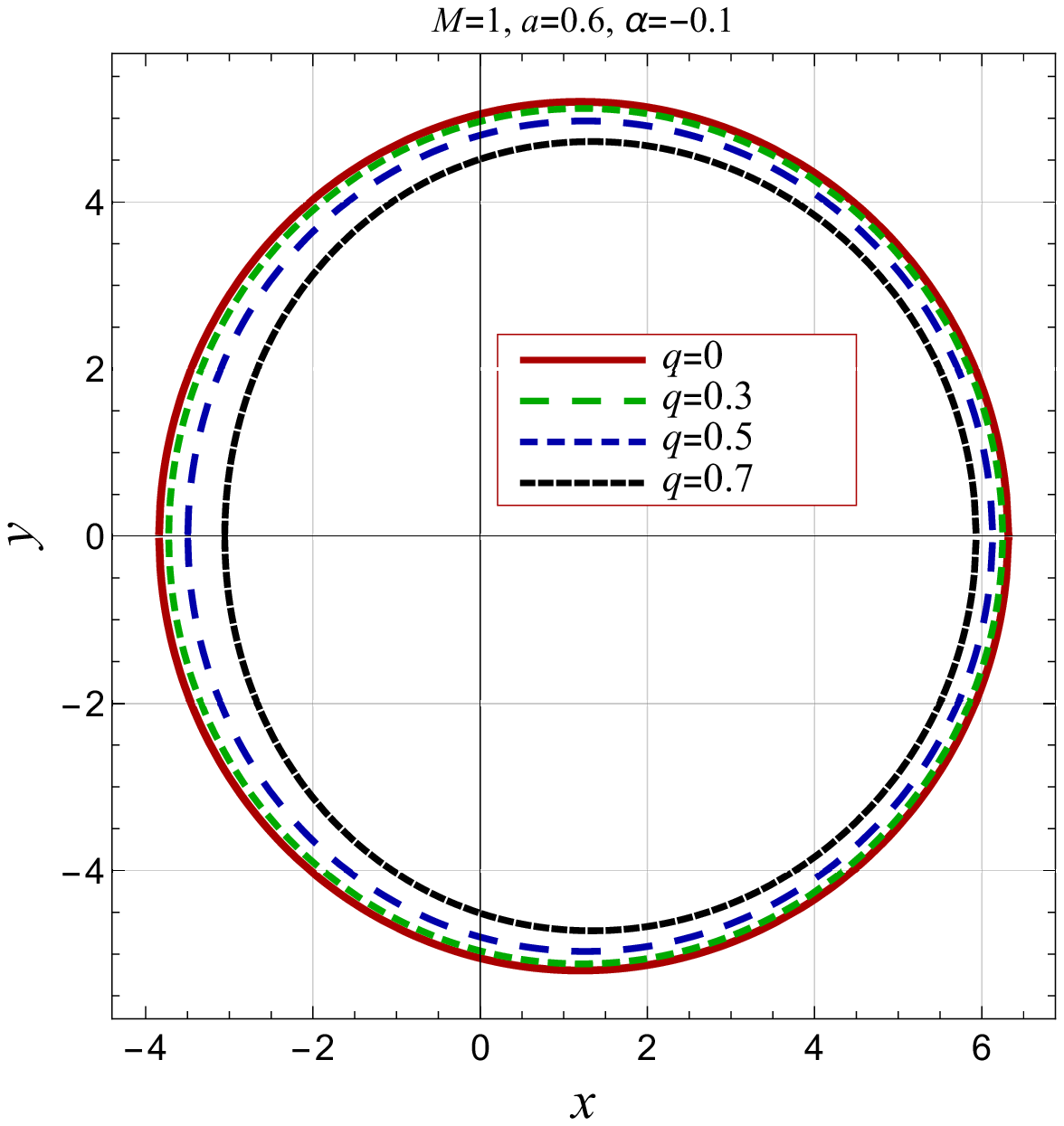}
\end{tabular}
\caption{\label{spl1} Plot showing the behavior of shape of the shadow of rotating charged black 
hole with Weyl corrections for different values of $a$ and $q$, for $\alpha (<0)$.}
\end{figure*}

\begin{figure*}[t]
\begin{tabular}{c c c c}
\includegraphics[scale=0.35]{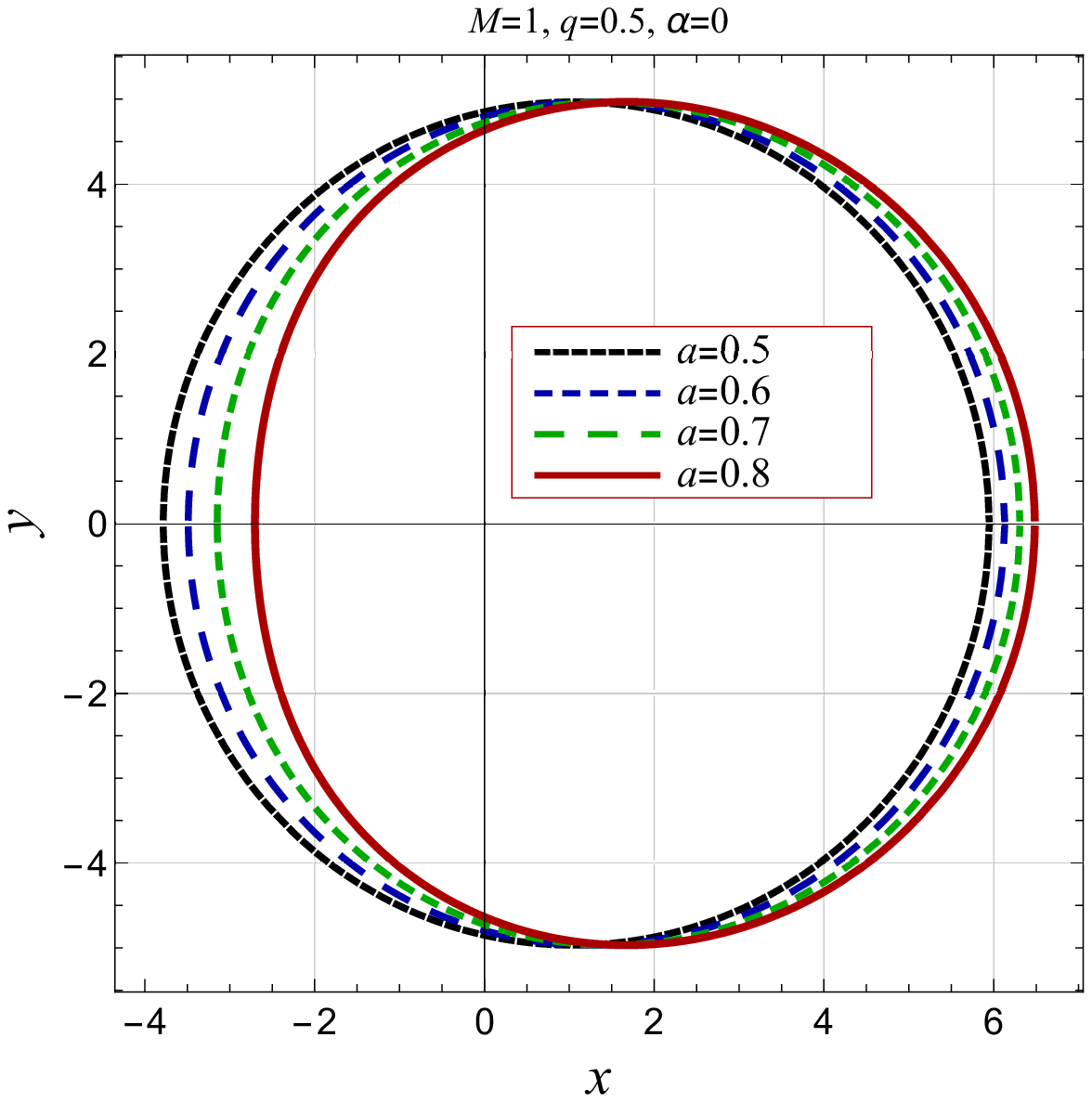}
\includegraphics[scale=0.35]{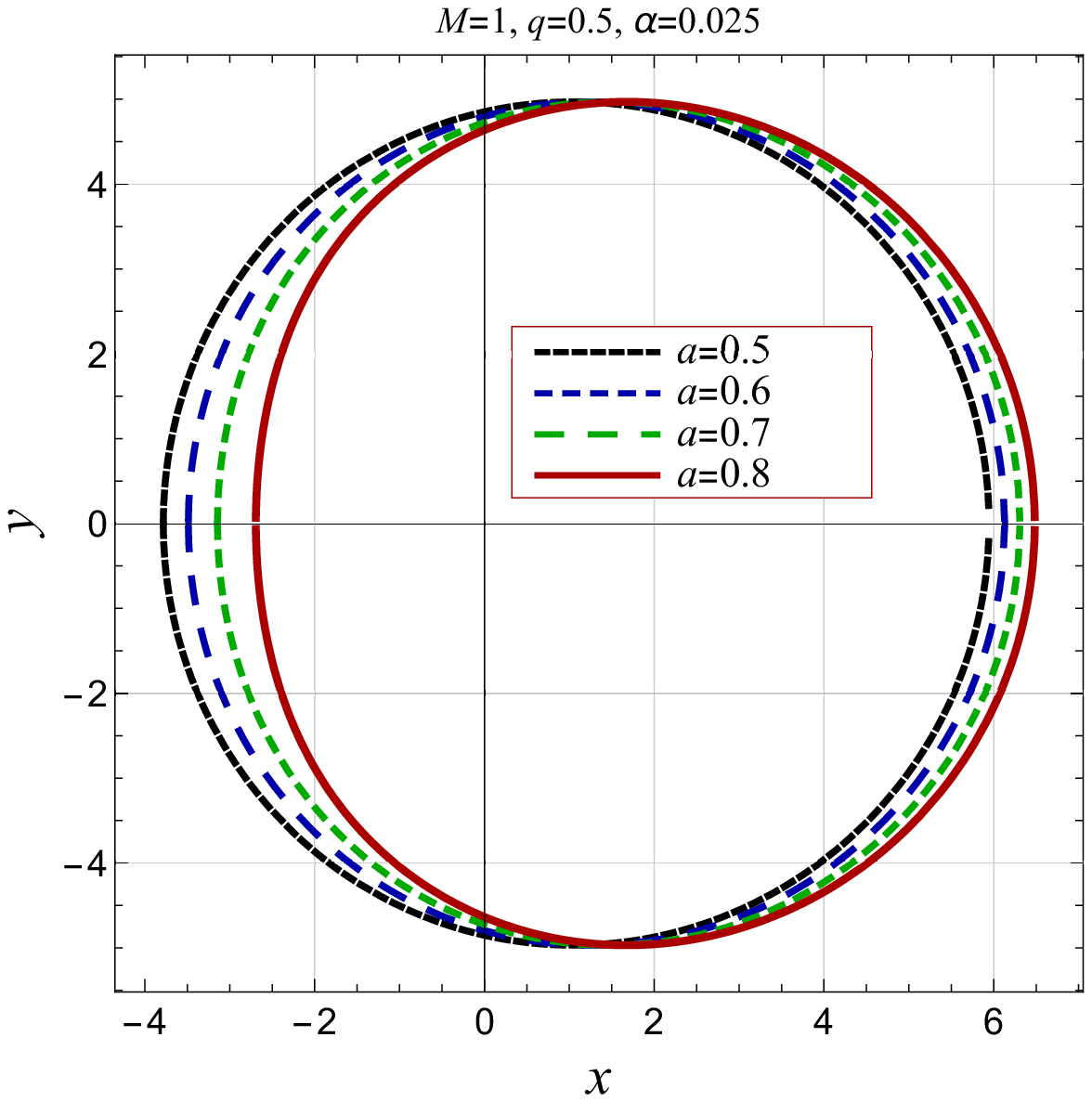}
\includegraphics[scale=0.35]{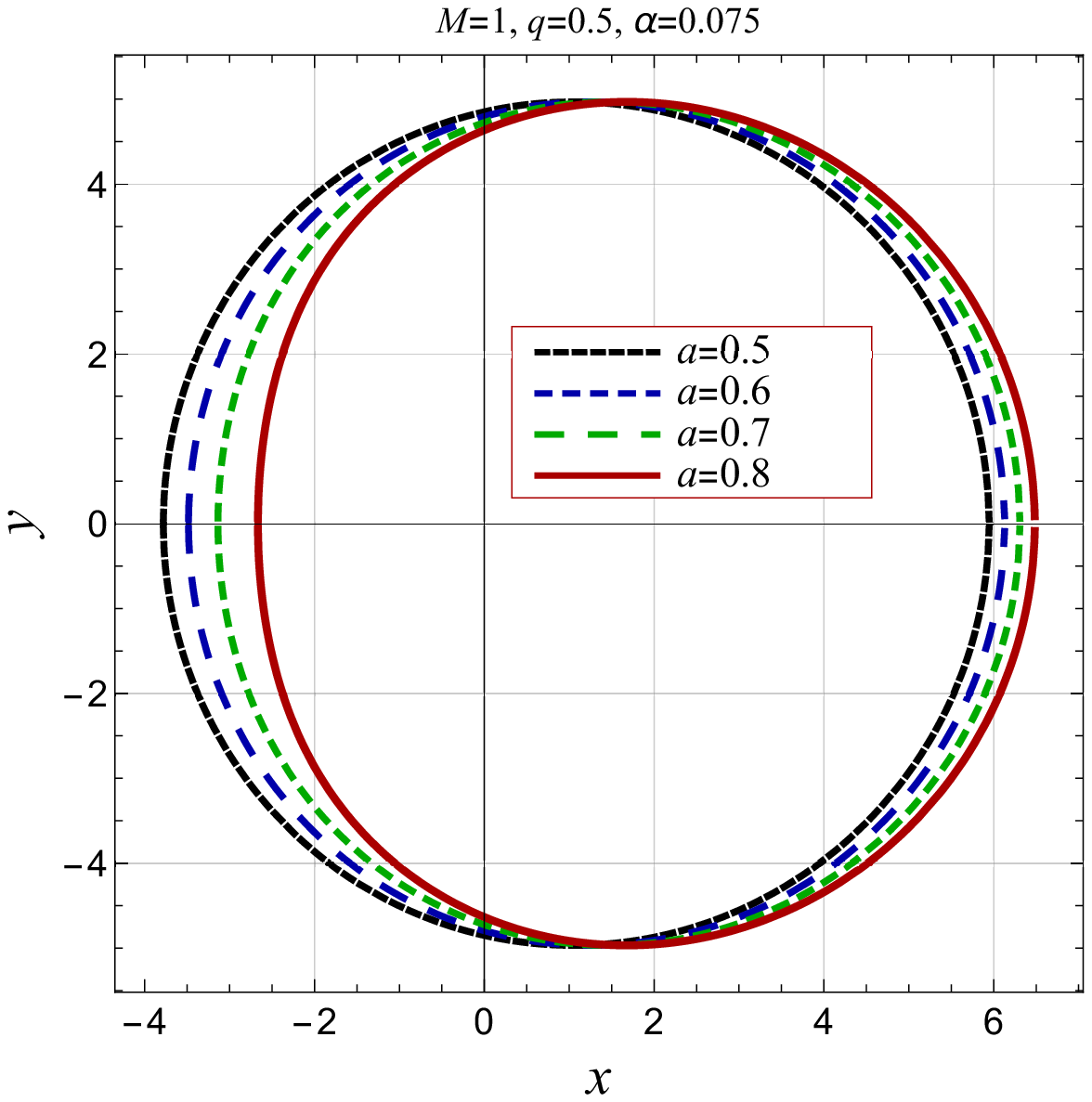}
\includegraphics[scale=0.35]{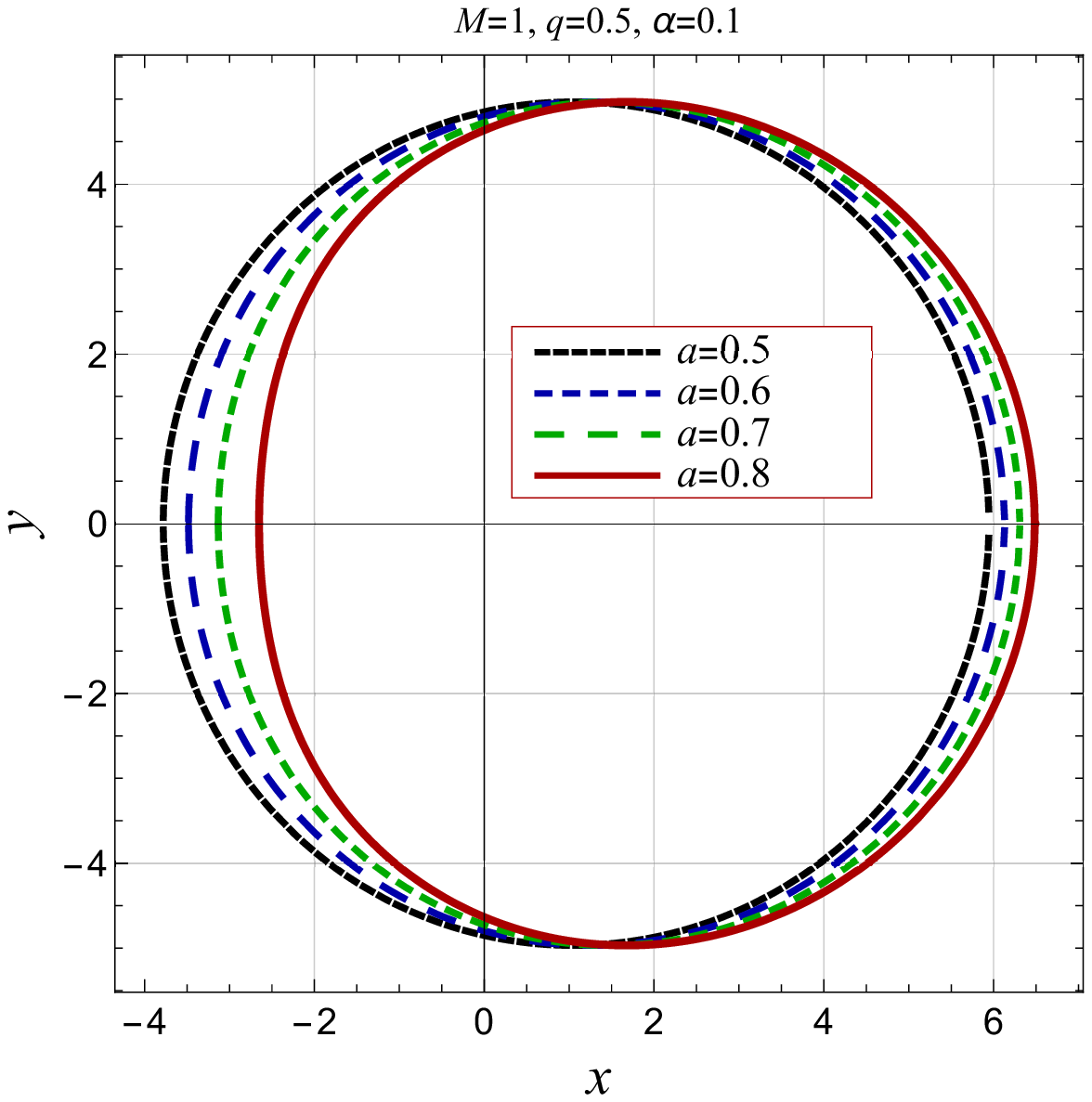}\\
\includegraphics[scale=0.35]{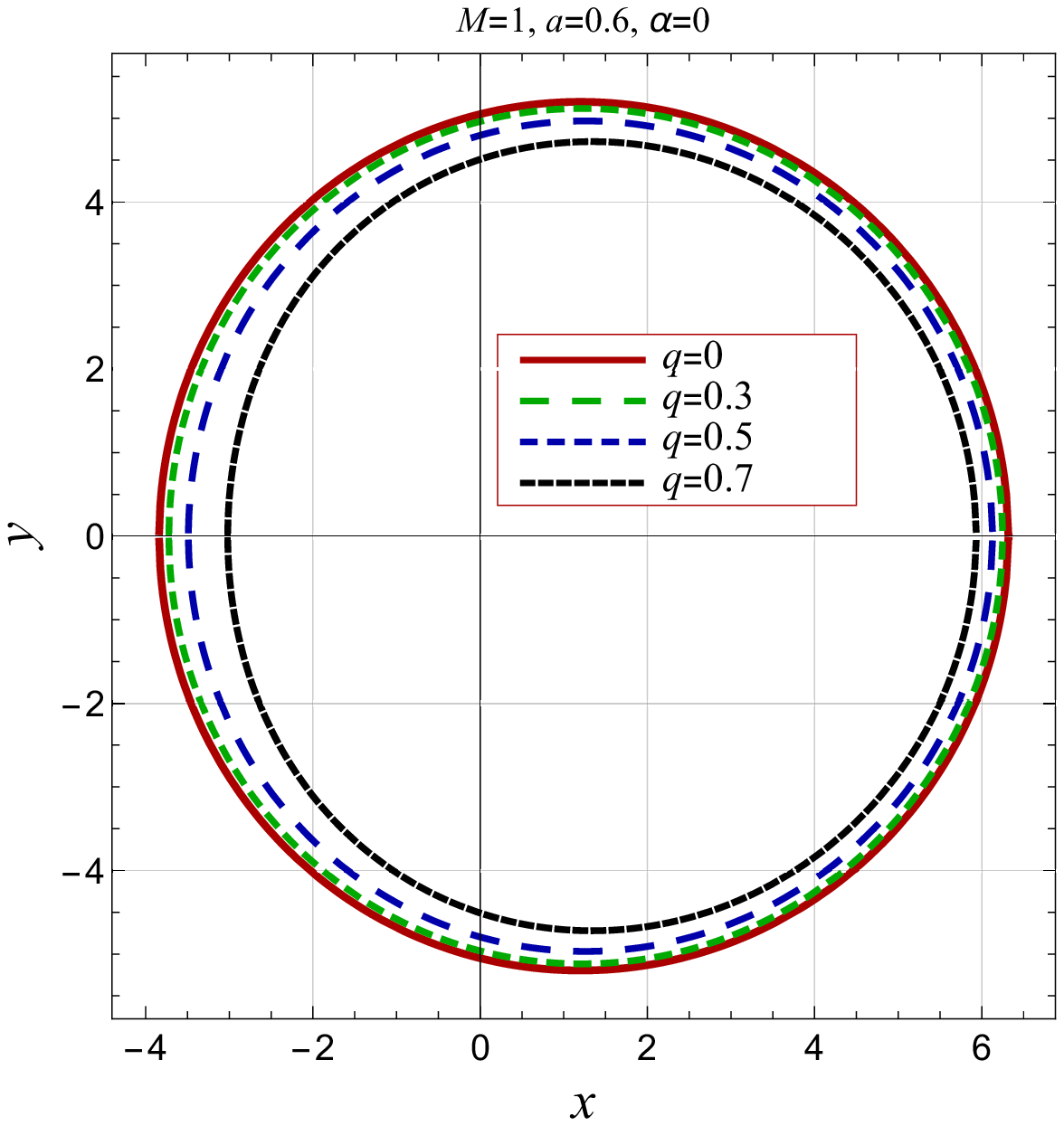}
\includegraphics[scale=0.35]{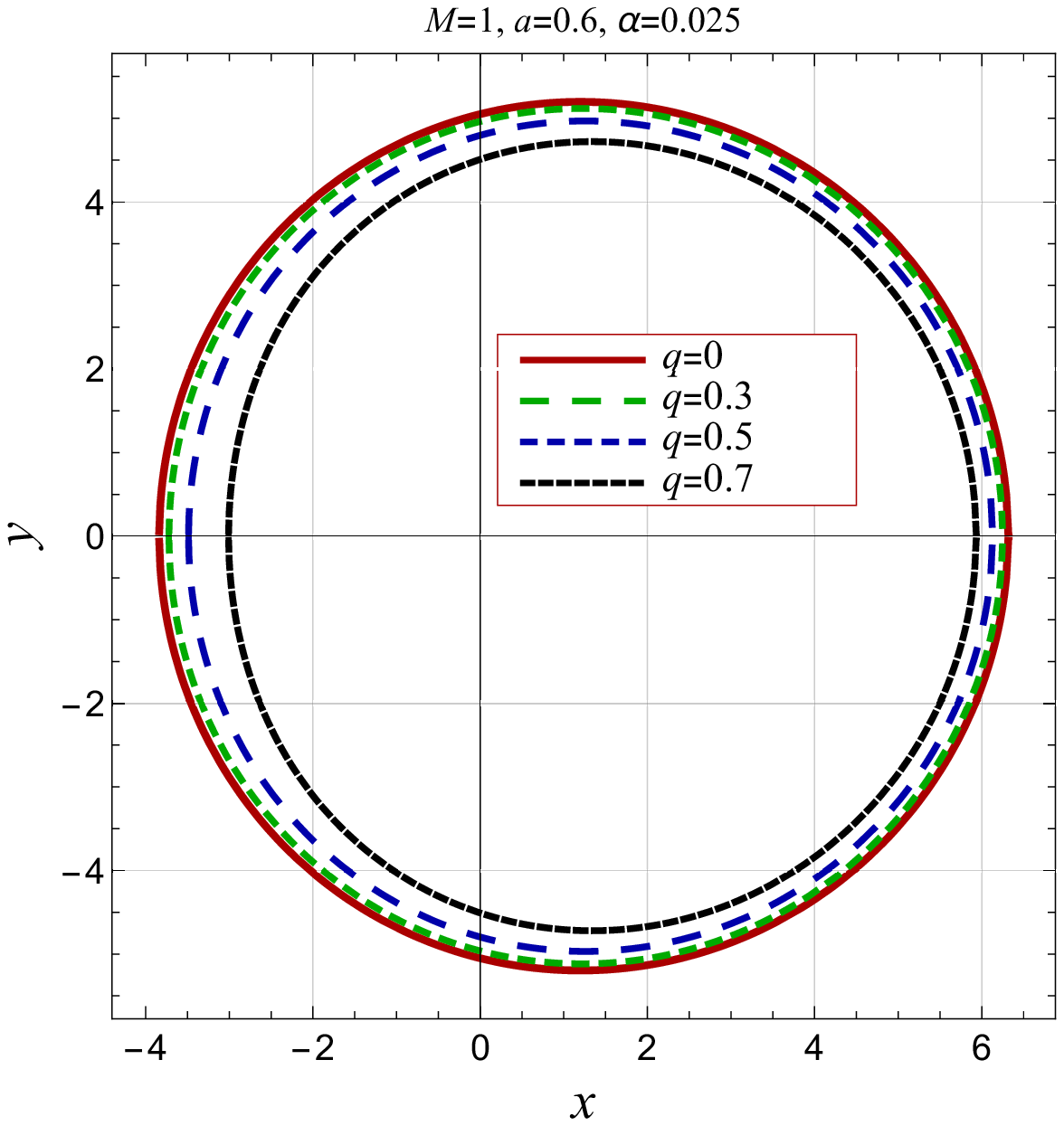}
\includegraphics[scale=0.35]{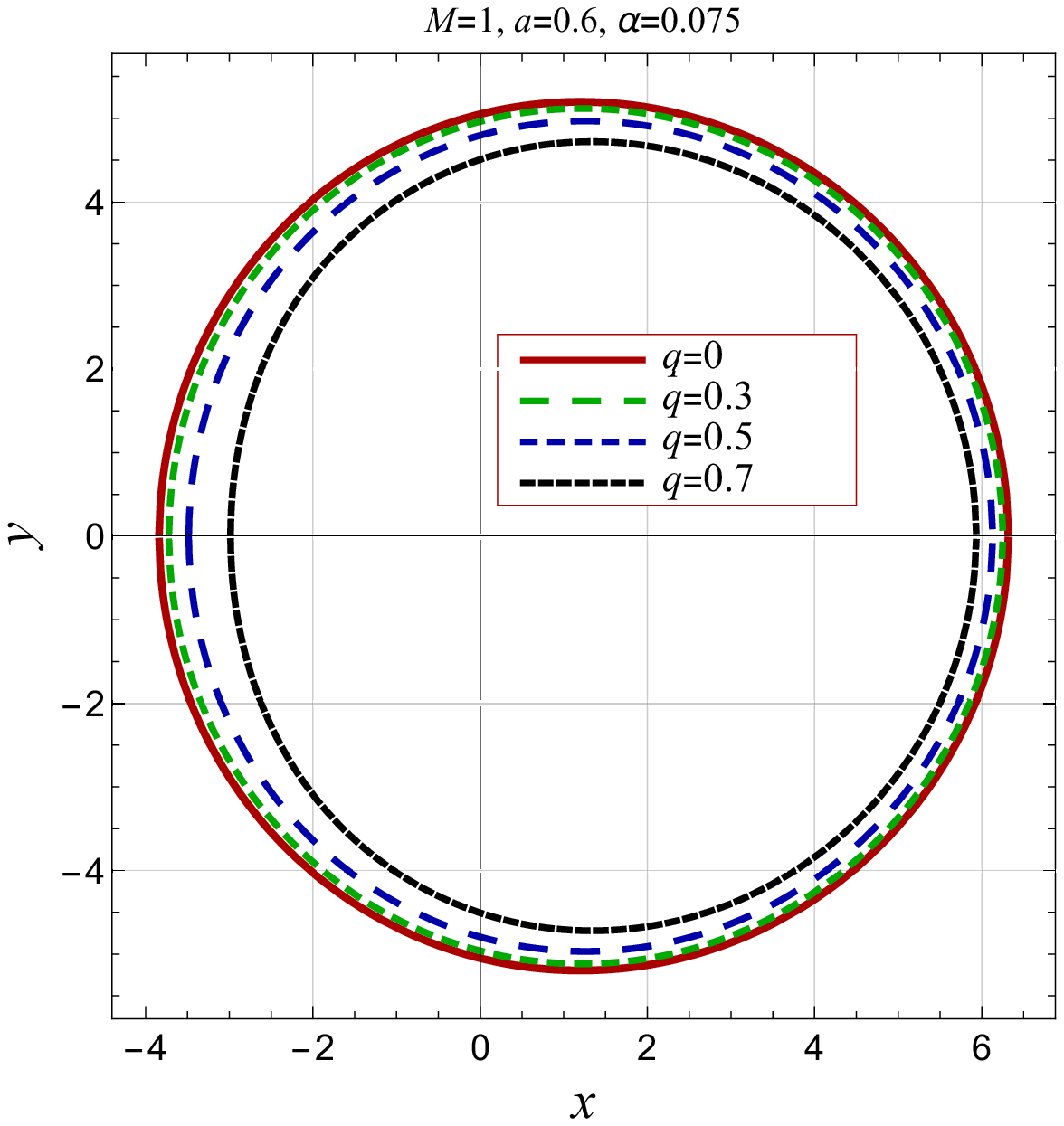}
\includegraphics[scale=0.35]{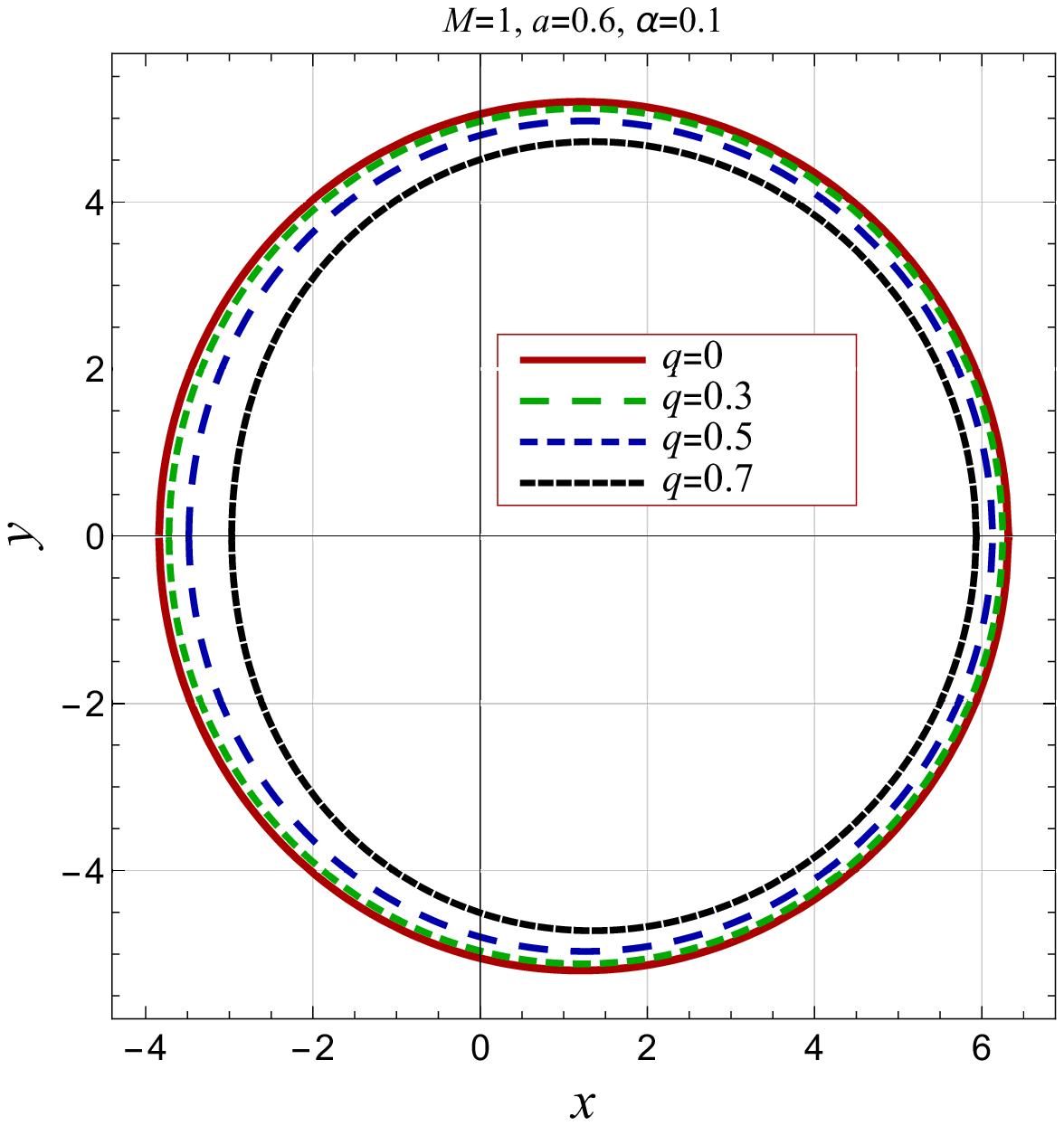}
\end{tabular}
\caption{\label{spl2} Plot showing the behavior of shape of the shadow of rotating charged black 
hole with Weyl corrections for different values of $a$ and and $q$, for $\alpha (>0)$.}
\end{figure*}

In the present section, we wish to construct the shapes of shadow for the Weyl corrected rotating charged 
black hole. The term black hole shadow can be understood as that the black hole is situated in between light source 
and a distant observer. A bunch of emitted photons gets deflection due to the strong gravitational field of the 
black hole. Although the photons with small impact parameter fall into the black hole as a result form a dark 
region in observer's sky. The distant observer which is observing this dark region would observe it as a dark spot on the 
bright background of light source. This dark region or dark spot is recognized as the \emph{shadow of black hole}. 
Note that the radius of black hole shadow is always larger than that of the geometrical size of event horizon. The apparent 
shape of black hole for a distant observer can be determined with the construction of celestial coordinates ($x$, $y$) 
\cite{Bardeen:1973gb,Chandrasekhar:1992}, which can be defined as follows
\begin{eqnarray}\label{celes}
x &=& \lim_{r_0 \rightarrow \infty}\left(-r_0^2\sin\theta_0\frac{d\varphi}{dr}\right), \nonumber\\
y &=& \lim_{r_0 \rightarrow \infty}\left(r_0^2 \frac{d\epsilon}{dr}\right),
\end{eqnarray}
where $\theta_0$ is the angle of inclination between the rotation axis of the black hole and observer's 
line of sight. The celestial coordinates $x$ and $y$ represent the apparent perpendicular 
distances of the image around the black hole and it can be seen from axis of symmetry and from its 
projection on the equatorial plane (cf. Fig.~\ref{celest}). On using Eqs.~(\ref{tphi}), (\ref{rtheta}), 
(\ref{RT}), and (\ref{celes}), we can get a relationship in between celestial coordinates and impact parameters
\begin{eqnarray}
x = -\xi, \quad y = \pm\sqrt{\eta}.
\end{eqnarray}
These expressions are very important from the point of view to construct the shapes of black hole shadow. 
We plot these equations for certain combination and variation of the parameters $\alpha$, $q$, and $a$ 
(cf. Fig.~\ref{spl1} and \ref{spl2}). It turns out that an asymmetry arises in the shape of black hole shadow. 
This asymmetry arises because of the rotation parameter $a$ and increases for large values. We construct different 
shapes of the black hole shadow by choosing both negative and positive values of coupling constant $\alpha$ 
(cf. Fig.~\ref{spl1} and \ref{spl2}). In order to demonstrate the deviations in the shape of black hole shadow, 
we include the Kerr and Kerr-Newman cases in the figures. Note that $\alpha=0$ represents the Kerr-Newman spacetime 
while $\alpha=q=0$ denotes the Kerr spacetime.

\begin{figure*}[t]
\begin{tabular}{c c c c}
\includegraphics[scale=0.55]{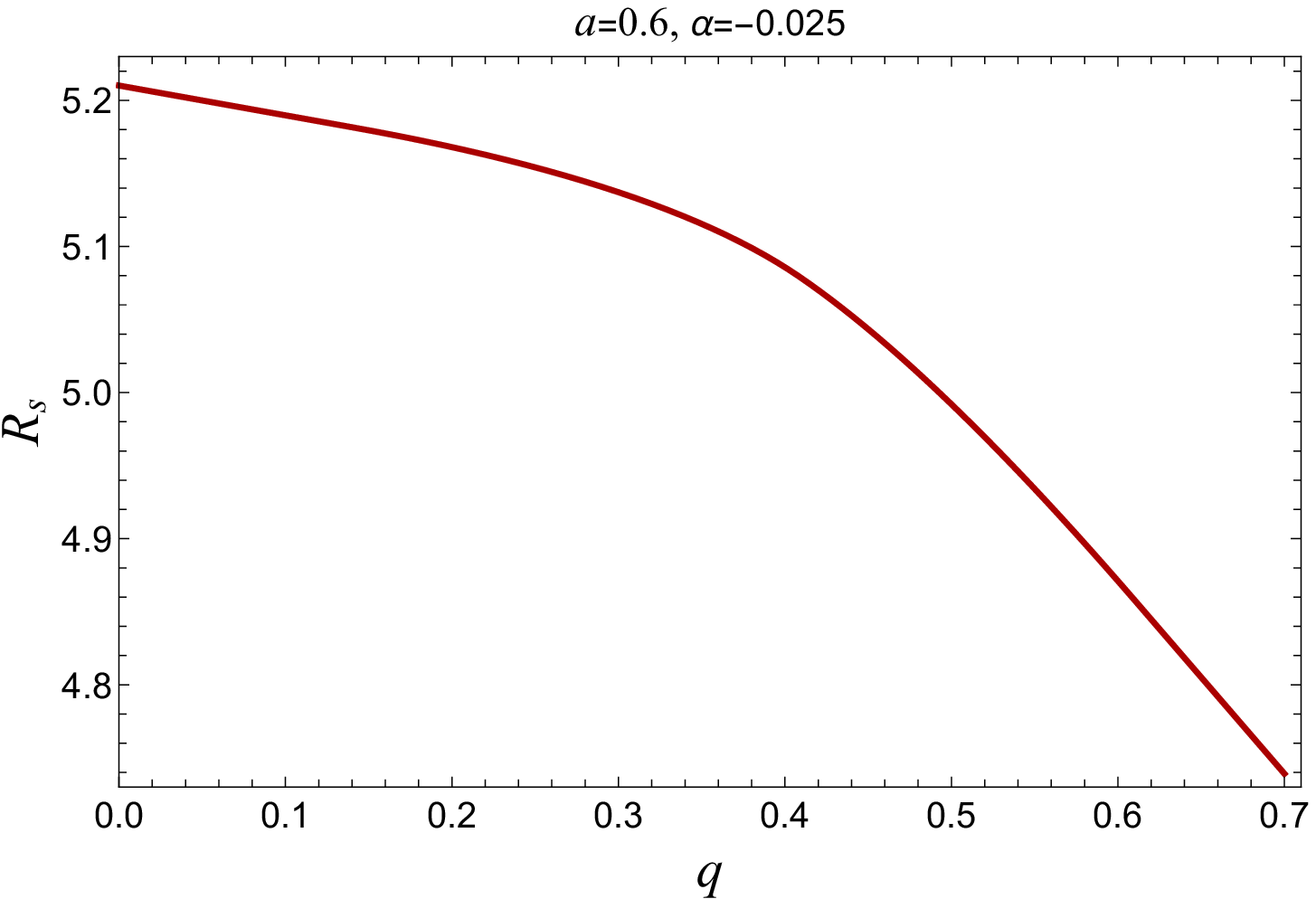}
\includegraphics[scale=0.55]{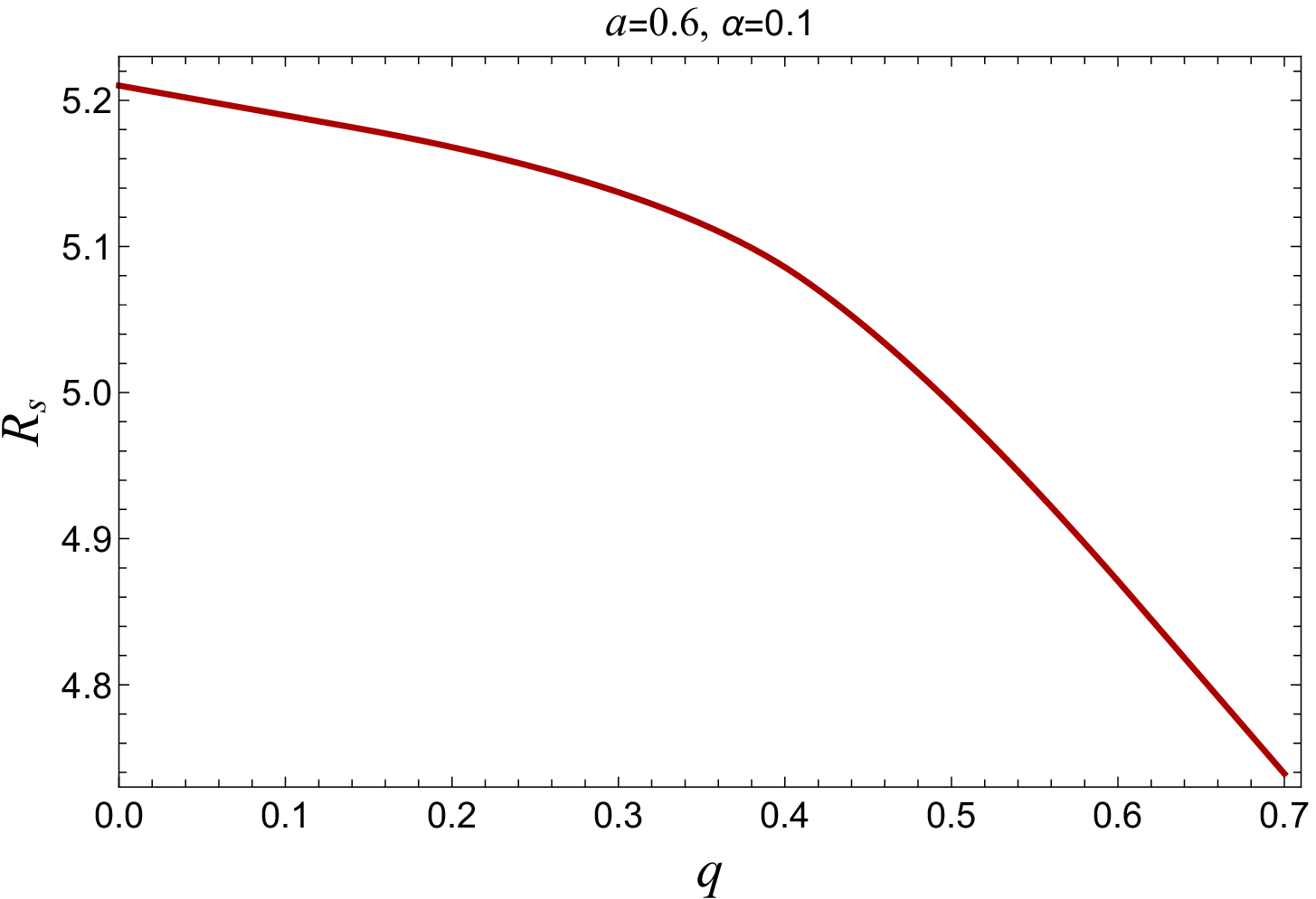}\\
\includegraphics[scale=0.55]{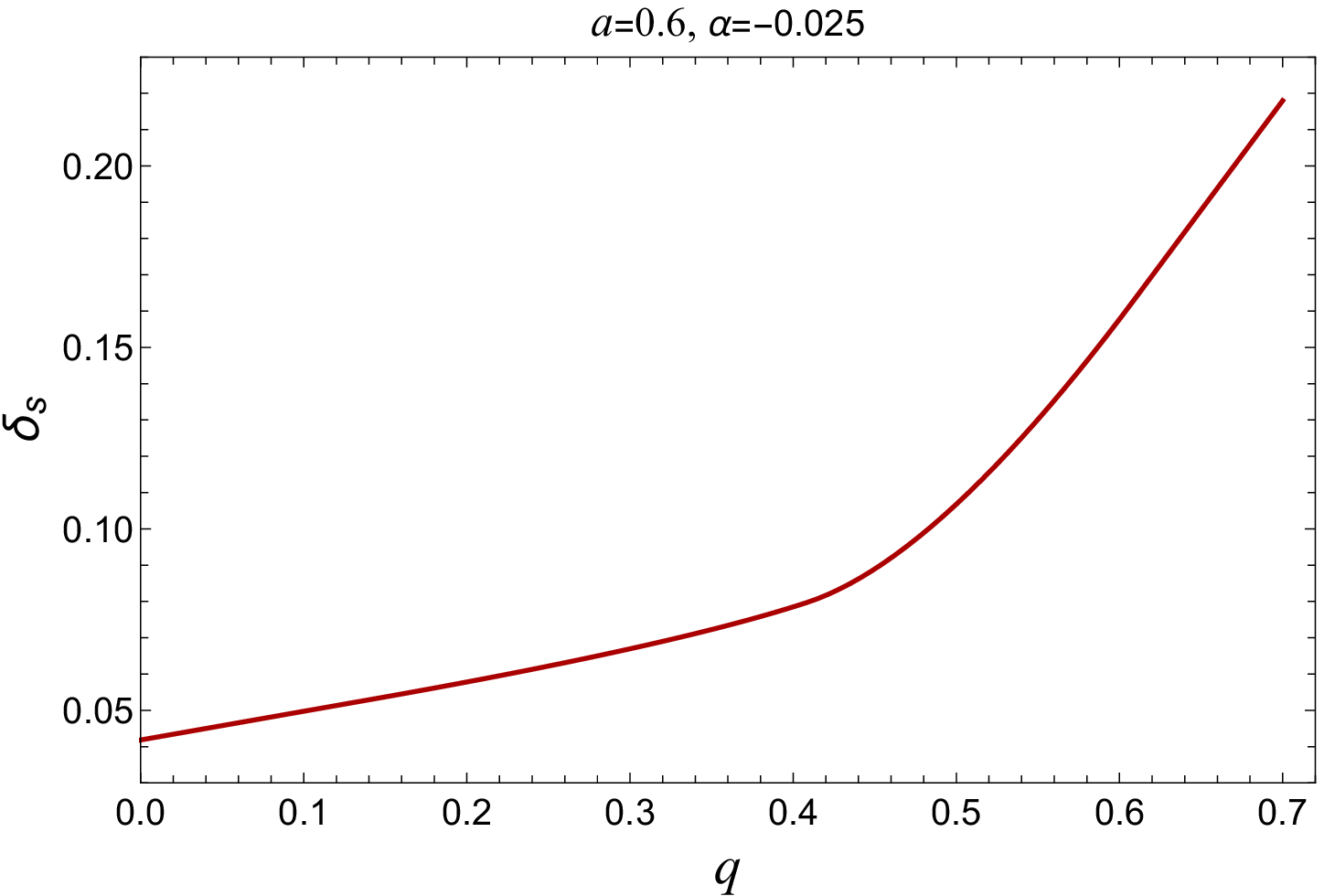}
\includegraphics[scale=0.55]{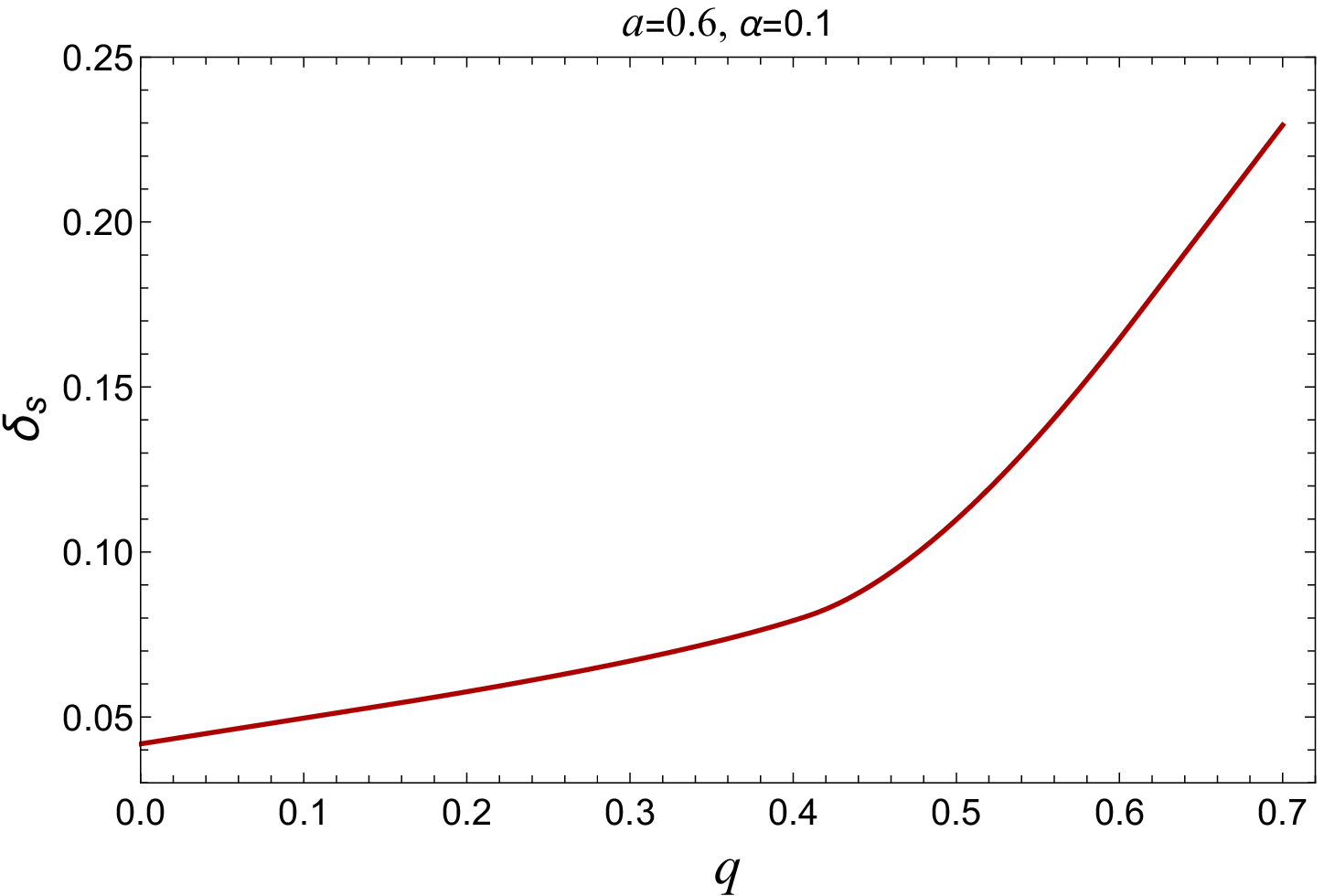}
\end{tabular}
\caption{\label{rsds} Plot showing the behavior of observables of the shadow for different values of both 
positive and negative values of $\alpha$.}
\end{figure*}

Moreover, let us have a look on the observables, namely radius $R_s$ and distortion $\delta_s$, which are 
responsible to give rise significant information about the black hole shadow. We are going to use the most popular 
definition of the observables given by Hioki and Maeda \cite{Hioki:2009na}:
\begin{eqnarray}
R_s &=& \frac{(x_t - x_r)^2 +y_t^2}{2 |x_t - x_r|}, \nonumber\\
\delta_s &=& \frac{(\tilde{x_p} - x_p)}{R_s},
\end{eqnarray}
where $(\tilde{x_p},0)$ and $(x_p,0)$ are the points where the reference circle and the 
silhouette of the shadow cut the horizontal axis at the opposite side of $(x_r,0)$ \cite{Hioki:2009na}. 
These observables are plotted in Fig.~\ref{rsds}; where the effect of the parameters $\alpha$ and $q$ on 
the shadow is depicted. It is found that the radius of the black hole shadow decreases with electric charge 
$q$ for both negative and positive values of the coupling constant $\alpha$. Besides, the distortion in the 
shape of the black hole shadow increases with $q$ in both of the cases. The presence of three free parameters in 
the spacetime renders more images of the shadow and open a gate to measure these parameters. It maybe possible once 
we will have the direct image of black hole so that we will be able to extract information regarding these parameters 
and test the gravity theory as well.

\section{Shadow in the presence of plasma}
\label{plasma}
In this section, we turn our attention in the construction of the geodesics equation of motion in the the plasma environment. Previous studies tells that the presence of 
plasma in surroundings of black holes affects the equations of motion of photon. It turns out that  
the influence of plasma may affect the shapes of black hole shadow. Throughout the calculations, we 
assume the plasma to be unmagnetized. The electron plasma frequency is given by
\begin{equation}
\omega_{e}^2(r)=\frac{4\pi e^2}{m_e} N(r),
\end{equation}
where $N(r)$ represents the electron number density in plasma. The notations $e$ and $m_{e}$ are 
stand for the electron charge and mass, respectively. As discussed in \cite{Synge:1960}, the refractive index of the plasma measured by 
a static observer is given in the following form
\begin{equation}
n^2 = 1+\frac{p^{\alpha}p_{\alpha}}{\left(p_{\beta}u^{\beta}\right)^2},
\end{equation} 
where $p^{\alpha}$ and $u^{\alpha}$ represents the four-momentum and four-velocity, respectively.
Therefore effective energy of the photon as measured by the distant observer has the form 
 $\hbar\omega=-p_{\alpha}u^{\alpha}$. Since in the absence of plasma the refractive index must have 
the value, $n=1$. We are going to use the two frequencies of the photons: one of them associated 
with the timelike Killing vector $t^{\alpha}$, i.e., $\omega_{t} \equiv 
-k^{\alpha}t_{\alpha}$ and the 
other one is associated with the four-velocity $u^{\alpha}$ of the distant static observer, i.e., 
$\omega \equiv -k^{\alpha}u_{\alpha}$, where $k^{\alpha}$ is a null wave vector \cite{Atamurotov:2015nra}. 
The refractive index $n$ of this plasma depends on the electron frequency in the plasma and the 
frequency of the photon as measured by a distant observer is given as
\begin{equation}
n^2=1-\frac{\omega_e^2}{\omega^2},
\end{equation}
where $\omega$ is the photon frequency, measured by a distant observer having four-velocity 
$u^{\alpha}$. Now we consider a radial power law density $N(r) = N_0/r^h$ where $h \geq 0$. Thus the 
electron plasma frequency takes the form: $\omega_e^2/\omega^2 = k/r^h$ where $k$ is a constant and $h$ is another constant which is
set to be 1. It is clear that the $k$ has a direct dependency with plasma frequency. 

We know that the Hamiltonian for the photon 
motion around the black hole gets modified in the presence of the plasma 
\cite{Atamurotov:2015nra}. It takes the following form for an arbitrary black hole 
\begin{eqnarray}\label{hamilton}
H \left(x^\alpha,p_{\alpha}\right) &=& \frac{1}{2}\left[g^{\alpha\beta}p_{\alpha} p_{\beta} 
+\left(n^2-1\right) \left(p_{\beta} u^{\beta}\right)^2\right] \nonumber\\
&=& \frac{1}{2}\left[g^{\alpha\beta}p_{\alpha} p_{\beta} +\hbar^2\omega_{e}^2\right].
\end{eqnarray}
Accordingly the equations of motions are also modified in the presence of plasma when calculated 
for the rotating charged black hole with Weyl corrections. The Hamiltonian looks similar to the 
Hamiltonian of massive particle in the vacuum. Note that we are interested in photon motion around 
the Weyl corrected rotating charged black hole. The Hamilton's equations of motion are written as 
follows
\begin{equation}
\frac{d x^{\alpha}}{d\sigma}=\frac{\partial H}{\partial p_\alpha},\quad 
\frac{{d p_\alpha}}{d\sigma}=-\frac{\partial H}{\partial x_\alpha}.
\end{equation}
As we already mentioned that the spacetime metric of rotating charged black hole with Weyl corrections contains two 
symmetries and the associated constants of motion are energy and angular momentum, which can be expressed by
\begin{equation}
E=-p_t=\hbar\omega, \quad L_z=p_\phi.
\end{equation}
Note that in the homogeneous medium, the plasma frequency $\omega_e$ is constant. Now we introduce 
the the dimensionless quantities as follows
\begin{equation}\label{hat}
\hat{E}=\frac{E}{\hbar\omega_e}=\frac{\omega}{\omega_e},\quad \hat{L_z}=\frac{L_z}{{\hbar\omega_e}}.
\end{equation}
On having all the necessary definitions and quantities in our hand, we are in a position to obtain the equations 
of motion for the photon in the presence of plasma. Now we use the Hamilton-Jacobi formulation to compute the 
equations of motion for the photon in equatorial plane, they take the following forms
\begin{eqnarray}
r^2 \lambda \frac{d{t}}{d{\sigma}} &=& \hbar\omega_e \left[a\left(\hat{L_z}-n^2a\hat{E}\right)
+\frac{\mathcal{T}}{\Delta} \left(r^2+a^2+\frac{4\alpha q^2}{9r^2} \right) \right],\nonumber\\
r^2 \lambda \frac{d{\varphi}}{d{\sigma}}&=& \hbar\omega_e 
\left[\left(\hat{L_z}-a\hat{E}\right)+\frac{a}{\Delta}\mathcal{T} \right], \nonumber\\
r^2 \lambda \frac{d{r}}{d{\sigma}} &=& \pm \hbar \omega_e \sqrt{\mathcal{R}_{pl}}, \nonumber\\
r^2 \lambda \frac{d \epsilon}{d\sigma} &=& \pm \sqrt{ \Theta_{pl}},
\end{eqnarray} 
where $\sigma$ is an affine parameter along the geodesics, 
$\mathcal{T} := \left(r^2+a^2+\frac{4\alpha q^2}{9r^2}\right)n^2\hat{E}-a\hat{L_z}$, 
$\mathcal{R}_{pl}$ and $\Theta_{pl}$ are expressed as follows
\begin{eqnarray}\label{Rpl}
\mathcal{R}_{pl} &=& \mathcal{T}^2 +\left(r^2+a^2+\frac{4\alpha q^2}{9r^2}\right)^2
\left(n^2-1\right)\hat{E}^2 \nonumber\\ 
&& -\Delta \left[\mathcal{\hat{K}}+\left(\hat{L_z}-a\hat{E}\right)^2\right],  \nonumber\\
\Theta_{pl} &=& \mathcal{\hat{K}}, \quad \hat{\mathcal{K}}=\mathcal{K}/\hbar\omega_e.
\end{eqnarray}
Since the conditions for the unstable spherical photon orbits are given by
\begin{equation}\label{ucond}
\mathcal{R}_{pl}=0, \quad \frac{d\mathcal{R}_{pl}}{dr}=0. 
\end{equation}
On inserting $\mathcal{R}_{pl}$ from Eq.~(\ref{Rpl}) into Eq.~(\ref{ucond}), and after some straightforward calculation, 
we arrive at the expressions of the impact parameters $\eta$ and $\xi$, which have the following forms
\begin{eqnarray}
\xi &=& \frac{\mathcal{B}}{\mathcal{A}}+\sqrt{\frac{\mathcal{B}^2}{\mathcal{A}^2}
-\frac{\mathcal{C}}{\mathcal{A}}}, \nonumber\\
\eta &=& \frac{\left(\psi-a\xi\right)^2+\psi^2\left(n^2-1\right)}{\Delta}-\left(\xi-a\right)^2,
\end{eqnarray}
where $\psi:=\left(r^2+a^2+\frac{4\alpha q^2}{9r^2}\right)$, the other notations $\mathcal{A}$, 
$\mathcal{B}$, and $\mathcal{C}$ are defined as follows
\begin{eqnarray}
\mathcal{A}&=&\frac{a^2}{\Delta},\nonumber\\
\mathcal{B}&=&\frac{a}{\Delta}\frac{\psi\Delta^{\prime}-\psi^{\prime}\Delta}{\Delta^{\prime}(r)}
\nonumber\\
\mathcal{C}&=&n^2\frac{\psi^2}{\Delta}+2\frac{\psi\psi^{\prime}n^2
-\psi^2n n^{\prime}}{\Delta^{\prime}}.
\end{eqnarray}
The prime $(')$ appears in the above expression is stand for the derivative with respect to radial 
coordinate $r$.

Consequently, the celestial coordinates of the distant observer in observers' sky are modified in 
the presence of plasma as follows 
\begin{eqnarray}\label{celplsma}
x &=& -\frac{\xi}{n}, \nonumber \\
y &=& \frac{\sqrt{\eta + a^2 -n^2 a^2}}{n}.
\end{eqnarray}
The boundary of the shadow can be determined with the help of Eq.~(\ref{celplsma}), which reflect an effect of plasma on 
the shape of black hole shadow.

\section{Conclusion}
\label{concl}
This paper aimed to develop the theoretical investigation of the black hole shadow for rotating charged 
black hole with Weyl corrections. We prospect that this investigation will be relevant and useful for the 
future observations by the Event Horizon Telescope. The black hole spacetime describes coupling in between 
electromagnetic and gravitational field. This spacetime is important from the astrophysical point of view 
because the recent investigations explain such coupling could exist near highly massive astrophysical 
objects.

We have started by briefly reviewing the spacetime metric with Weyl corrections, aimed to determine all the 
relevant expressions necessary to investigate the black hole shadow. Thus we have derived the photon 
geodesics around the black hole spacetime by using the Hamilton-Jacobi method. We further computed the 
expressions of the impact parameters on employing the spherical photon orbits condition. In order to 
construct images of the black hole shadow, we have introduced celestial coordinates as well as determined a 
relationship between the observer's celestial coordinates and impact parameters. Indeed, these expressions 
are essential to construct the apparent image or shadow of the black hole in the celestial plane. We have 
demonstrated how the presence of Weyl corrections in the spacetime affects the shape and size 
of the black hole shadow. In order to show it graphically, we have plotted different cases of black hole 
shadow by choosing the different set of values of the parameters, namely $\alpha$, $a$, and $q$. We have
discovered that the shapes of black hole shadow has an asymmetry due to the presence of rotation parameter 
$a$ as well as charge $q$ and coupling constant $\alpha$. It has been noticed that for large values of 
rotation parameter $a$, the shape of the black hole shadow distorted more in both cases of positive and 
negative $\alpha$. In further investigation, the influence of charge $q$ on the radius of black hole shadow 
has been explained which implies that the radius of black hole shadow decreases for both positive and 
negative values of coupling constant $\alpha$. Besides, we have also explored the black hole shadow of 
rotating charged black hole with Weyl corrections in the presence of unmagnetized plasma. The necessary 
formulation to describe black hole shadow has been developed in the plasma environment. We have discovered 
that the black hole shadow has been affected by surrounding plasma environment.   

\acknowledgments
 M.A. would like to thank University of KwaZulu-Natal and the National Research Foundation for financial support. 
\end{document}